\documentclass[titlepage,12pt]{utarticle}
\usepackage{amsmath,amssymb,amsthm,amscd,cite}
\usepackage{latexsym,amsmath,amssymb,graphicx}
\usepackage[all]{xy} 
\usepackage{epsf}
\xyoption{v2}
\xyoption{2cell}
\xyoption{dvips}
\CompileMatrices
\LaTeXdiagrams
\UseAllTwocells
\numberwithin{equation}{section}
\newcommand{\ra}{\rightarrow}
\newcommand{\be}{\begin{equation}}
\newcommand{\ee}{\end{equation}}
\newcommand{\refeq}[1]{(\ref{eq:#1})}
\newcommand{\IP}{\mathbb{P}}
\newcommand{\co}{{\cal O}}

\newcommand\IZ{\mathbb {Z}}
\newcommand{\IC}{\mathbb{C}}
\newcommand{\IE}{\mathbb{E}}
\newcommand{\IR}{\mathbb{R}}
\newcommand{\ba}{\begin{array}}
\newcommand{\ea}{\end{array}}
\newcommand{\BX}{\overline{X}}
\newcommand{\BT}{\field{T}}
\newcommand{\bu}{\overline{u}}
\newcommand{\bv}{\overline{v}}
\newcommand{\bz}{\overline{z}} 
\newcommand{\bw}{\overline{w}}
\newcommand{\om}{\overline{M}}

\newcommand{\lb}{\lambda} 
 
\begin{document}
\preprint{
  RUNHETC-2003-13\\
  HU-EP-03/21\\
  {\tt hep-th/0305021}\\
}
\title{
Orientifolds, Unoriented Instantons and  Localization}
\author{Duiliu-Emanuel Diaconescu$^{\sharp}$, Bogdan Florea$^{\sharp}$ and Aalok Misra$^{\natural}$}
\oneaddress{
      \smallskip
      {\centerline {\it $^{\sharp}$ New High Energy Theory Center, Rutgers University,}}
      \smallskip 
      {\centerline {\it Piscataway, NJ 08854-0849, USA}}
      \smallskip
      {\centerline {\it $^{\natural}$ Institut f\"ur Physik, Humboldt Universit\"at zu Berlin,}}
      \smallskip
      {\centerline {\it Invaliden Stra\ss e 110, D-10115, Germany}}
}
\date{May 2003}

\Abstract{
We consider world-sheet instanton effects in ${\cal N}=1$ 
string orientifolds of noncompact 
toric Calabi-Yau threefolds. 
We show that unoriented closed string topological 
amplitudes can be exactly computed using localization techniques 
for holomorphic maps with involution.
Our results are in precise agreement with mirror symmetry and large 
$N$ duality predictions. 
}

\maketitle 

\section{Introduction}\label{sec:intro}

It is widely known that oriented closed string instantons play a central role 
in the ${\cal N}=2$ dynamics of Calabi-Yau compactifications. 
In particular one can obtain exact 
results for the prepotential and more general couplings by counting (in an 
appropriate sense) holomorphic maps from the world-sheet to the target space. 

Similar results have been obtained in the past years for ${\cal N}=1$ D-brane 
vacua, in which case the nonperturbative effects are due to open string 
instantons 
\cite{AV:mirror,AKV:disk,DVi:matrix,DVii:geom,GJS,GZ,KL,LM,LMWi,LMWii,LS,Mi,
Mii,OV:knot}.
Although this case is more intricate from a mathematical 
point of view, open string enumerative techniques have been developed 
up to the point of concrete computations. From a physical point of view 
these effects often generate a nonperturbative superpotential, with the 
effect of removing the vacuum degeneracy of the models in question. 

Another class of ${\cal N}=1$ string vacua which has been less studied so far 
consists of orientifolds of Calabi-Yau threefolds equipped with an 
antiholomorphic involution \cite{AAHV,BH,AM1,AM2,SV,VW}. 
Even in the absence of D-branes, these models 
exhibit interesting nonperturbative effects due to closed string 
unoriented instantons. So far exact results for topological unoriented string 
amplitudes on the resolved conifold have been predicted in \cite{SV} 
as a result of large $N$ duality. 
Moreover, the $\BR\BP^2$ result has been checked using local mirror symmetry 
in 
\cite{AAHV}. 

The purpose of the present paper is to develop a systematic approach 
to this problem based on an unoriented version of closed string 
enumerative geometry. 
More precisely 
we propose a method of summing unoriented world-sheet instantons 
based on localization of holomorphic maps with involution. 
Our approach is similar to the open string {\bf A}-model techniques 
of \cite{GZ,KL,LS} which rely on localization with respect to a circle action. 
This is not an entirely rigorous construction since one directly sums over 
the fixed loci in the absence of a rigorous 
intersection theory on an appropriate moduli 
space along the lines of \cite{B,BF,GP,K,KM,LT}. 
In very concrete terms the computation 
reduces to a sum over Kontsevich graphs with involution. 
Nevertheless, we will show that the results obtained by this technique 
are in perfect agreement with large $N$ duality and mirror symmetry 
predictions. The present approach is valid  
for all genera and in principle it does not require a large $N$ dual 
description. 
We illustrate this method for a concrete toric model obtaining a 
generating functional which satisfies the 
integrality properties predicted in \cite{SV}. 
We also compare our results for $\IR\IP^2$ instanton effects 
to a {\bf B}-model local mirror computation obtaining precise 
agreement. 

The paper is structured as follows. In section one, we 
review some basic facts about 
topological orientifolds and discuss the basic principles 
of the method. Section three is devoted to concrete computations 
for a local conifold geometry, and in section four we treat a more 
involved toric model.

\section{Orientifolds of Topological {\bf A}-Models} 

We begin our analysis with some preliminary remarks on orientifolds 
of closed string topological {\bf A}-models using 
\cite{Hi,Hii,Hiii} as basic references. In the following we will elaborate 
on instanton effects in the unoriented sector of the theory. 

Let $X$ be a toric noncompact Calabi-Yau threefold equipped with an 
antiholomorphic involution $I:X\ra X$. Throughout this paper we will 
only consider freely acting involutions. 
The orientifold models are obtained  
by gauging a discrete symmetry of the form 
$\sigma I$ where $\sigma: \Sigma 
\ra \Sigma$, therefore the orientifold group 
is $G=\{1,\sigma I\}\simeq \IZ/2$.  
If $\Sigma$ has genus zero, the action of $\sigma$ is given by 
by $w\ra -{1/{\bar w}}$ in terms of an affine coordinate $w$. 
The action of $\sigma$ on surfaces of higher genus $g$ is more intricate. 
The classification of antiholomorphic involutions of genus $g$ 
surfaces goes back to Felix Klein \cite{FKi,FKii}. 
It is known that these involutions 
are completely characterized by three invariants $(g,n(\sigma),k(\sigma))$
where $n(\sigma)$ is the number of connected components of the fixed
locus $\Sigma_\sigma$ 
of $\sigma$, and $k(\sigma)$ is the index of orientability. 
The latter is defined to be two minus the number of connected components of 
$\Sigma\setminus \Sigma_\sigma$. 
Here we are interested in involutions of type $(g,0,1)$. 
In this case $\sigma$ is an orientation reversing diffeomorphism of $\Sigma$
so that the quotient $\Sigma/\langle \sigma \rangle$ is an unoriented 
surface without boundary. For future reference we will denote the 
cyclic group of order two $\langle \sigma \rangle$ by $G_{ws}$. 
Note that not all Riemann surfaces of a given genus admit such involutions. 
Those surfaces that admit antiholomorphic involutions are usually 
called symmetric Riemann surfaces. It is known that 
symmetric surfaces of type $(g,0,1)$ exist for any $g\geq 0$. 

In order to construct the partition function of the theory, one has to sum 
over twisted sectors \cite{Hi,Hiii}, just as in ordinary orbifold theories.
For a space-time orbifold theory (in which the action on the world-sheet 
is trivial) we have the well known formula 
\be\label{eq:orbA}
Z_g = \frac{1}{|G|^g} 
\sum_{\alpha\in {\rm Hom}(\pi_1(\Sigma),G)} Z_g(\alpha).
\ee
The sum in the right hand side of \eqref{eq:orbA} is over all representations 
of the fundamental group of $\Sigma$ in the orbifold group $G$. 
In orientifold theories, we have a similar formula except that 
the fundamental group of $\Sigma$ must be replaced by the 
orbifold fundamental group $\pi_1O(\Sigma)$ defined with respect 
to the action of the world-sheet group $G_{ws}$. $\pi_1O(\Sigma)$
is the group of all diffeomorphisms of the universal cover 
${\widetilde \Sigma}$ of $\Sigma$ which are lifts of elements of $G_{ws}$. 
For concrete applications, note that there is an exact sequence of the form 
\be\label{eq:exseqA} 
1\ra \pi_1(\Sigma) \ra \pi_1O(\Sigma) \ra G_{ws} \ra 1.
\ee
Then the twisted sectors of the theory are defined by representations 
$\alpha \in {\rm Hom}(\pi_1O(\Sigma),G)$ \cite{Hiii}
such that we have a commutative 
triangle diagram 
\be
\xymatrixrowsep{2pc}
 \xymatrixcolsep{2pc}
\begin{diagram}
\pi_1O(\Sigma) \ar[r]^\alpha \ar[d]& G \ar[dl]\\
G_{ws}. & \\
\end{diagram}
\ee
The vertical arrow in the above diagram is the projection 
$\pi_1O(\Sigma) \ra G_{ws}$ of equation (\ref{eq:orbA}) with kernel  
$\pi_1(\Sigma)\subset \pi_1O(\Sigma)$. Therefore 
the monodromy $\alpha$ is always trivial along the 
generators of $\pi_1(\Sigma)$. 
The map $G\ra G_{ws}$ is defined by $\sigma I \ra \sigma$. 
This means that there is only one twisted sector of the theory 
consisting of maps $f:\Sigma\ra X$ satisfying the equivariance 
condition 
\be\label{eq:equivconA}
f\circ \sigma = I \circ f. 
\ee

So far these considerations are not specific to topological sigma 
models. In a topological theory, the partition function reduces to 
an instanton sum. The twisted sector instantons 
are area minimizing maps $f:\Sigma\ra X$ satisfying the equivariance 
condition (\ref{eq:equivconA}). 
Here the area must be defined with respect to a 
K\"ahler metric $g$ on $X$ left invariant by the involution $I:X\ra X$. 
If $g$ is an arbitrary K\"ahler metric on $X$, then $(g+I^*g)/2$ is 
an invariant metric in the same K\"ahler class. Since we are studying a 
topological {\bf A}-model, the invariant metric need not be Calabi-Yau. 

The equivariance condition (\ref{eq:equivconA}) implies that $f$ descends 
to a map between quotients 
${\widetilde f}:\Sigma/G_{ws} \ra X/I$. Since the involutions $\sigma, I$ 
reverse orientation, the quotient spaces $\Sigma/G_{ws}, X/I$ do not have 
complex analytic structures. Instead they carry natural {\it dianalytic} 
structures \cite{Alling}. Without giving too many details, a dianalytic 
structure is defined by an atlas (up to equivalence) whose transition 
functions on overlaps are either holomorphic or antiholomorphic. 
Then ${\widetilde f}$ is a dianalytic function if it is either 
holomorphic or antiholomorphic on each connected component of the domain. 
This means that in the unoriented theory 
we will have to sum over both holomorphic and antiholomorphic instantons. 
Since $\Sigma/G_{ws}$ is connected, the two sums will be identical and 
the net effect is an overall factor of 2. Keeping this in mind, it suffices to 
consider only equivariant holomorphic maps. 

In order to compute virtual instanton numbers we need to set up an 
intersection problem on an appropriate moduli space. From the previous 
paragraph, it follows that the moduli space in question should classify 
equivariant holomorphic maps up to a certain equivalence relation. 
Two such maps $(\Sigma, f , \sigma)$, $(\Sigma',f', \sigma')$ 
are said to be equivalent if there exists an isomorphism $\phi:\Sigma \ra 
\Sigma'$ compatible with the antiholomorphic involutions so that 
$f=f'\circ \phi$. The compatibility condition for $\phi$ is very similar 
to \eqref{eq:equivconA} 
\be\label{eq:equivconB}
\phi \circ \sigma = \sigma'\circ \phi.
\ee 
In principle one should give a rigorous construction of this moduli space 
and then define a virtual cycle of degree zero which counts the 
virtual instanton numbers. This program has been carried out for 
oriented holomorphic closed string maps in the context 
of Gromov-Witten theory \cite{B,BF,GP,K,KM,LT}.
Loosely speaking, in the present context one should 
construct an unoriented version of this theory. 

However, this is not the route we 
will take in the present paper. Instead we will employ a semirigorous  
approach inspired from a similar treatment of open string Gromov-Witten 
theory \cite{GZ,KL,LS}. To recall the basic points, note that in the open 
string framework one would like to count in appropriate sense holomorpic 
maps defined on bordered Riemann surfaces with boundary conditions 
specified by a lagrangian 3-cycle $L\subset X$. Although the moduli 
space of such maps is not rigorously understood, one can take a direct 
approach to this problem in the presence of a real torus action on $X$ 
preserving $L$. The main point is that it is much easier to handle the 
fixed loci of the induced action on the moduli space instead of the 
moduli space itself. In particular one can give a computational definition 
of the virtual cycle using the localization formula of Graber and 
Pandharipande \cite{GP} adapted to
the open string context. The main ingredient of this computation is the 
deformation complex of open string maps. One peculiar aspect 
of this approach is that the resulting virtual instanton numbers 
depend on the choice of toric weights. This dependence reflects 
the presence of a (real codimension one) 
boundary in the moduli space of maps, and it is partially 
understood in the context of large $N$ duality. 
More details on this subject can be found in \cite{KL}. 

Our proposal is that a similar approach can be implemented for
equivariant holomorphic maps as well. To begin with, we need a real
torus action on $X$ which is compatible with the antiholomorphic 
involution $I$. Since $X$ is a toric noncompact threefold, it admits  
a $T^3$ action. Here we want to find a one dimensional subtorus 
$T\subset T^3$ 
which preserves the antiholomorphic involution $I$. Moreover, in order 
for the localization approach to be effective, the action of $T$ 
should have only isolated fixed points on $X$. These conditions 
are somewhat restrictive, but we will show that such good actions 
exist in specific examples. 
Granting the existence of $T$, there is an induced action on the 
moduli space of equivariant holomorphic maps. The next step is 
to enumerate all fixed loci of the induced action. Then one can assign 
a local contribution to each connected component of the fixed locus
using an equivariant version of the localization theorem of \cite{GP}. 
This approach is very similar to the original work of Kontsevich 
on localization and Gromov-Witten invariants \cite{K}. We will show 
in the next section that the fixed loci are naturally classified 
in terms of Kontsevich graphs with involution. 
In order to compute the instanton numbers, one has to sum 
the local contributions over all fixed loci. Since we have not developed 
the fundamental theory, this procedure involves some subtle sign ambiguities 
which can be fixed by additional arguments. Essentially, one has to keep in
mind that the final answer has to be a rational number independent 
of toric weights, if the moduli space is properly compactified. 
Here we will use a compactification of the moduli space inspired from
\cite{K,KM}. Namely we will allow the domain symmetric Riemann surfaces
$\Sigma$ to develop ordinary double points so that the resulting nodal 
surfaces admit freely acting antiholomorphic involutions. 
We must also impose a stability condition which makes the automorphism 
group of a symmetric map finite. Then we will sum over all fixed loci of the 
torus action satisfying these conditions, and the result should be a 
rational number. Note that there is an important difference between 
the case considered here and the open string maps studied in \cite{KL}. 
In the latter case, the compactified moduli space turns out to be a 
space with a real codimension one boundary. The boundary is associated 
to degenerations of the domain bordered surface in which nodes appear 
on boundary components. This is clearly a real codimension one 
phenomenon. In our case, such phenomena are absent because we consider 
only freely acting involutions, therefore the are no real codimension 
one fixed loci and no real codimension one phenomena. 
All these ideas will be made more concrete below. 

\section{Localization and Invariant Graphs I} 

In order to illustrate the basic principles we will first carry out the 
above program for a resolved conifold geometry. Exact results for 
topological unoriented amplitudes have been predicted in \cite{SV} 
using large $N$ duality. 
Here we will show how these results can be reproduced by pure 
{\bf A}-model computations. This example has also been considered in
\cite{AAHV} in the context of mirror symmetry. 

The toric threefold 
$X$ is isomorphic to the total space of the rank two bundle $\co(-1)
\oplus \co(-1)$ over $\IP^1$, which is given by the following toric data 
\be\label{eq:toricA} 
\begin{array}{ccccc}
& X_1 & X_2 & X_3 & X_4 \cr
\IC^* & 1 & 1 & -1 & -1.\cr
\end{array}
\ee
The second homology of $X$ is generated by the zero section $C\subset X$, 
which is an isolated $(-1,-1)$ rational curve. Moreover, this is the 
only projective curve on $X$. 

The freely acting antiholomorphic involution considered in \cite{AAHV}
is given by 
\be\label{eq:ahinvolA}
I:\left(X_1,X_2,X_3,X_4\right)\ra \left(\BX_2,-\BX_1,\BX_4,-\BX_3\right).
\ee
Note that the quotient $X/I$ is a smooth unorientable manifold which carries 
a dianalytic structure. The involution \eqref{eq:ahinvolA} preserves the zero 
section, therefore $C/I$ is a dianalytic 
two-cycle on $X/I$ with topology $\BR\BP^2$. 
This cycle generates the second homology of $X/I$. 
For future reference we introduce local coordinates 
$z=\frac{X_1}{X_2}, u={X_2X_3}, v={X_2X_4}$ on the patch 
$\{X_2\neq 0\}\subset X$. 
Then the involution \refeq{ahinvolA} reads $\left(z,u,v)\ra(-\frac{1}{ 
\bz}, -\bz\bv, \bz\bu\right)$. 

As explained in the previous section, counting unoriented 
world-sheet instantons reduces to counting 
holomorphic maps $f:\Sigma \ra X$ subject to the 
equivariance condition $f=I\circ f \circ \sigma$. 
Therefore at this point it may be helpful to recall some basic facts 
about the enumerative geometry of $X$. 
The main ingredients of Gromov-Witten theory can be found in \cite{KL}:

$a)$ a compact moduli space 
$\om_{g,0}(X,\beta)$ of stable maps to $X$ with fixed genus $g$ 
and fixed homology class $\beta =f_*[\Sigma]\in H_2(X)$,

$b)$ a virtual fundamental cycle $[\om_{g,0}(X,\beta)]^{vir}$ on the moduli 
space,

$c)$ an orientation on the moduli space. 

\noindent
If $X$ is a Calabi-Yau threefold, the virtual cycle has dimension
zero and the virtual number of maps is defined to be the degree 
of this cycle
\be\label{eq:GWa}
C_{g,\beta}=\int_{[\om_{g,0}(X,\beta)]^{vir}}1.
\ee
This data gives rise to a truncated Gromov-Witten potential 
of the form 
\be\label{eq:GWb} 
\CF_X=\sum_{g\geq 0}\sum_{\substack{\beta\in H_2(X) \\ \beta\neq 0}}
C_{g,\beta}q^\beta
\ee
where $q^{\beta}$ is a formal symbol satisfying $q^{\beta+\beta'}=
q^\beta q^{\beta'}$. 

In the present case, any holomorphic map $f:\Sigma \ra X$ 
must factorize according to the following diagram 
\be\label{eq:factA}
\xymatrixrowsep{2pc}
 \xymatrixcolsep{2pc}
\begin{diagram}
\Sigma \ar[r]^f \ar[dr]& X \\
& C. \ar@{^{(}->}[u]\\
\end{diagram}
\ee
This means that the moduli space of stable 
holomorphic maps $\om_{g,0}(X,d[C])$ is isomorphic to 
the moduli space of stable degree $d$ maps to $\IP^1$, $\om_{g,0}(\IP^1,d)$.
The virtual fundamental cycle can be constructed using a perfect obstruction
complex on the moduli space \cite{BF,LT}. Moreover, there is a canonical
orientation induced by the holomorphic structure. 

The integral in the right hand side of (\ref{eq:GWa}) can be evaluated 
using the localization theorem of Graber and Pandharipande \cite{GP}. 
There is a $T=S^1$ action on $X$ given by 
\be\label{eq:toractA} 
e^{i\phi}\cdot(X_1,X_2,X_3,X_4)=(e^{i\lb_1\phi}X_1,e^{i\lb_2\phi}X_2,
e^{i\lb_3\phi}X_3, e^{i\lb_4\phi}X_4)
\ee
which induces a $T$ action on the moduli space $\om_{g,0}(\IP^1,d)$. 

Then one has a localization formula of the form 
\be\label{eq:locA}
C_{g,d}=\sum_{\Xi} \int_{[\Xi]^{vir}} {1\over e_{T}(N^{vir}_\Xi)}
\ee
where the sum is over all fixed loci of the torus action on the moduli 
space. In the right hand side $N^{vir}_\Xi$ is the virtual normal bundle 
to $\Xi$ and $[\Xi]^{vir}$ is the virtual cycle on $\Xi$ induced by the one 
on the ambient moduli space. $e_{T}$ denotes the equivariant Euler class. 
The contributions of the fixed loci can be evaluated using the 
tangent obstruction complex, as explained 
in \cite{GP,K}. 
Note that the individual terms in this sum take values in the fraction 
field of the representation ring of $T$, ${\cal R}_{T}$. Nevertheless the 
final result is a rational number as expected. 
Using these techniques, one can show that the Gromov-Witten potential 
\refeq{GWa} takes the following closed form \cite{FP}
\be\label{eq:GWc} 
\CF_X = \sum_{d\geq 1}{e^{-dt}\over d\left(2{\rm sin}{dg_s\over2}\right)^2}.
\ee

Here we have to solve a similar counting problem for 
holomorphic maps $f:\Sigma \ra X$ satisfying the extra condition 
$f = I\circ f \circ \sigma$. Ideally, one would like to follow 
steps $(a)-(c)$ outlined above, but this is not the 
route we will take here.  
Instead we will employ a less rigorous approach 
following similar ideas used in the context of 
open string enumerative geometry \cite{GZ,KL,LS}.  
The main point is that in practice the fixed loci of a torus action 
on the moduli space are much easier to describe than the moduli space 
itself. Moreover, the contribution of each fixed locus to the 
right hand side of \refeq{locA} can be easily evaluated using 
a local description of the perfect obstruction complex of the moduli 
space. Therefore, one could in principle obtain the final answer only 
from the data of the fixed loci supplemented with some local deformation 
theory. This is essentially the original approach proposed by Kontsevich 
in \cite{K} where the fixed loci are classified in terms of graphs. 
Then one can design a simple algorithm for computing the
contribution of a fixed locus in terms of the combinatorics of the 
associated graph. 

In order to implement this method in our situation, we need a 
torus action on $X$ compatible with the antiholomorphic involution. 
This will induce a torus action on the moduli space, and we can proceed
with the classification of all fixed loci. The local contribution of each 
fixed locus will be evaluated using an equivariant version of the local 
tangent-obstruction complex. Finally, the virtual number of equivariant 
holomorphic maps will be obtained by summing over all fixed loci.

There are a couple of subtle points here which should be discussed in some 
detail. In Gromov-Witten theory one obtains a compact moduli space by 
including degenerate maps $(\Sigma, f)$ subject to a stability condition. 
The domain $\Sigma$ is allowed to degenerate to a reducible singular curve 
with only ordinary double points. Stability requires the map $(\Sigma, f)$ 
to have a finite automorphism group. This means that any connected curve 
$\Sigma'\subset \Sigma$ which is mapped to a point must be stable 
in the sense of Deligne and Mumford \cite{DM}.

Our problem is slightly more 
complicated since we have to classify triples $(\Sigma, f,\sigma)$ where 
$\sigma$ is an antiholomorphic involution of $\Sigma$ of type $(g,0,1)$. 
For convenience, recall that two such triples are equivalent if there 
exists an isomorphism $\phi:\Sigma\longrightarrow \Sigma'$ such that 
$f=f'\circ\phi$ and $\sigma = \phi^{-1} \circ \sigma'\circ \phi$.
Therefore in order to obtain a compact moduli space, we have to include 
degenerate maps with involution. In particular, the domain $\Sigma$ of such
a map must be a symmetric nodal Riemann surface which admits a antiholomorphic 
involution of type $(g,0,1)$. 
The classification of such objects is not a simple task, 
so the structure of the moduli space is not easy to understand. 
However, for our purposes it suffices to understand only the 
structure of the fixed loci $\Xi$, which is a more tractable question. 
The main outcome is that such fixed loci are classified in terms 
of Kontsevich graphs with involution. Let us discuss these aspects in more 
detail for the resolved conifold geometry. 

Consider a generic torus action on $X$ of the form \eqref{eq:toractA}.
This action is compatible with the antiholomorphic involution 
\eqref{eq:ahinvolA} if the weights satisfy the conditions $\lb_1+\lb_2=0$, 
$\lb_3+\lb_4=0$. In terms of local coordinates $(z,u,v)$, the $T$ action 
reads 
\be\label{eq:toractC}
e^{i\phi}\cdot(z,u,v) = (e^{i\lb_z}z,e^{i\lb_u}u,e^{i\lb_v}v)
\ee
where $\lb_z=\lb_1-\lb_2$, $\lb_u=\lb_1+\lb_3$, $\lb_v=\lb_1+\lb_4$. 
The compatibility condition becomes 
\be\label{eq:toractD}
\lb_z+\lb_u+\lb_v=0. 
\ee
If this condition is satisfied, there is an induced 
torus action on the moduli space of holomorphic 
maps with involution. In the following we describe the structure of the 
fixed loci.

Let us first recall the Kontsevich graph representation of fixed loci 
in $\om_{g,0}(X,d[C])$ \cite{K}.
To any invariant map $f:\Sigma \ra X$ 
we can associate a connected graph 
$\Gamma$ as follows. Let $P_1,P_2$ denote the fixed points of the 
torus action on $X$ defined by 
\be\label{eq:fixedptsA}
P_1:\quad X_1=X_3=X_4=0,\qquad P_2:\quad X_2=X_3=X_4=0.
\ee
Note that both $P_1, P_2$ lie on $C$. 

i) The vertices $v\in V(\Gamma)$ represent connected components 
$\Sigma_v$ 
of $f^{-1}(P_1, P_2)$, which can be either points or disjoint unions 
of several irreducible components of $\Sigma$. To each vertex $v$ 
we associate a number $k_v\in {\{1,2\}}$ defined by $f(\Sigma_v)=P_{k_v}$, 
and the arithmetic genus $g_v$ of $\Sigma_v$.   

ii) The edges $e\in E(\Gamma)$ correspond to irreducible components 
of $\Sigma$ which are mapped onto $C$. By $T$-invariance, each such 
component $\Sigma_e$ must be a rational curve, and $f|_{\Sigma_e}:\Sigma_e
\ra C$ must be a Galois cover of degree $d_e\geq 1$. In terms of local 
coordinates the restriction of $f$ to $\Sigma_e$ is given by 
$z=w^{d_e}$. 

Then the claim of \cite{K} is that the set of all fixed loci is 
in one to one correspondence with (equivalence classes 
of) graphs $\Gamma$ subject to the following conditions 

1) If $e\in E(\Gamma)$ is an edge connecting two vertices $u, v$, 
then $f(u)\neq f(v)$. 

2) $1-\chi(\Gamma) +\sum_{v\in V(\Gamma)} g_v = g$, where $g_v$ 
is the arithmetic genus of the component $\Sigma_v$, and $\chi(\Gamma)$ 
is the Euler characteristic of $\Gamma$, 
$\chi(\Gamma) = |V(\Gamma)|-|E(\Gamma)|$. 

3) $\sum_{e\in E(\Gamma)} d_e = d$. 

For clarification, let us consider an invariant map of genus $g=4$ 
and degree $d=18$ as in fig. 1. Note that 
each horizontal component maps to $C$ with a certain 
degree as shown in the left figure. The corresponding marked graph  
is shown in the right figure. For each vertex $v$, 
the first integer represents the value of $f(v)$, i.e. the fixed point 
it maps to. The second integer is the arithmetic genus of $\Sigma_v$.

\begin{figure}[ht]
    \centering
    \scalebox{0.4}{\includegraphics[angle=0]{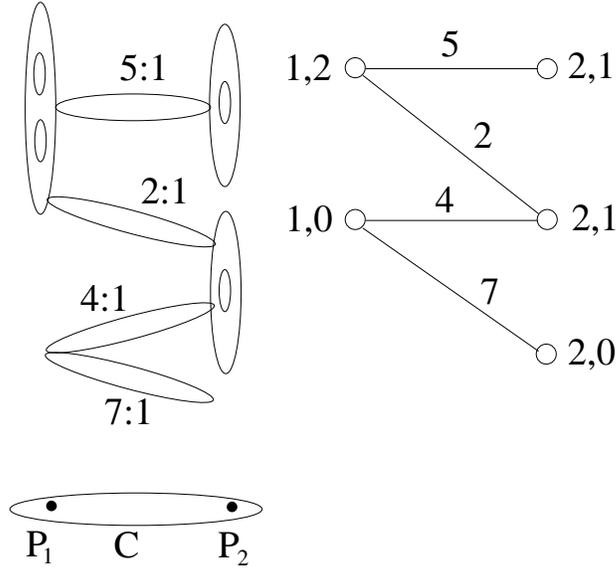}}
     \caption{The Kontsevich graph associated to a fixed map of degree 
18 and genus 4.}\label{graphsI}
\end{figure} 

Now let us consider stable holomorphic maps with involution. Recall that 
the involution $I$ acts on $C$ by $[X_1:X_2]\ra [\BX_2: -\BX_1]$, 
hence it exchanges 
the fixed points $P_1, P_2$. 
In local coordinates we have $z\ra -1/\bz$. 
The fixed loci consist of maps which satisfy 
$f=I\circ f \circ \sigma$. Therefore $\sigma$ must send
each component $\Sigma_v$ of the domain 
of type $(1,g_v)$ to another component of type $\Sigma_{v'}$ 
of type $(2,g_{v'})$. In particular the two components must have the 
same genus $g_v=g_{v'}$. 
This means that $\Sigma$ is symmetric under pairwise exchange of 
components $\Sigma_{v}$ of type $1$ and type $2$. The corresponding graphs 
must accordingly be symmetric under pairwise exchange of vertices of type one 
and two. 

The action of $\sigma$ on the horizontal components can be of two types.

$A)$ A component $\Sigma_e$ may be mapped to itself by $w\ra -1/\bw$, 
where $w$ is a local coordinate. Then the map $z=w^{d_e}$ 
is equivariant if and only if $d_e$ is odd. 

$B)$ A component $\Sigma$ may be mapped to another component $\Sigma_{e'}$
of the same degree $d_e=d_{e'}$. In terms of local coordinates $(w_1,w_2)$,
the involution must be given by $w_1\ra (-1)^{d_e-1}/{\bw_2}$, $w_2\ra 
(-1)^{d_e-1}/\bw_1$. Then one can easily check that the map $z=w_1^{d_e}$, 
$z=w_2^{d_e}$ is equivariant. 

In conclusion, the equivariance conditions impose important symmetry 
conditions on the fixed loci. These conditions are best summarized 
using the graph representation. The admissible graphs are characterized by 
the following conditions 

$(I)$ There is a involution $\tau_V:V(\Gamma)\ra V(\Gamma)$ 
sending a vertex of type one to a vertex of type two,
leaving the genus invariant $g_{\tau_V(v)}=g_v$. 

$(II)$ There is a involution 
 $\tau_E:E(\Gamma)\ra E(\Gamma)$ leaving the degree invariant, 
$d_{\tau_E(e)}=d_e$. If $e$ is a fixed point of $\tau_E$, $d_e$ must be 
odd. 

$(III)$ The involutions $\tau_E,\tau_V$ must be compatible in the following 
sense. If an edge $e$ is attached to a vertex $v$, then $\tau_E(e)$ must be 
attached to $\tau_V(v)$. 

Given such a graph $\Gamma$, we have to evaluate the local contribution of 
the associated fixed locus. Although we have not rigorously defined a virtual 
fundamental cycle, we will assume that such a construction exists and 
that a localization result of the type \cite{GP} holds in our case as well. 
Then the local contribution of a fixed locus can be computed 
using the tangent-obstruction complex of a holomorphic map with 
involution. 

Recall that the tangent-obstruction complex of an arbitrary holomorphic 
map $f:\Sigma \ra X$ is 
\be\label{eq:tangobsA}
\begin{aligned} 
0& \ra Aut(\Sigma)\ra H^0(\Sigma, f^*T_X) \ra {\BT}^1 
\ra Def(\Sigma) \ra H^1(\Sigma, f^*T_X) \ra {\BT}^2\ra 0.\cr
\end{aligned}
\ee
where ${\BT}^1$, ${\BT}^2$ are the deformation and respectively
obstruction space of $(\Sigma,f)$. 
This can be interpreted as a complex of sheaves on the moduli space. 
If $(\Sigma,f)$ represents a point of a fixed locus $\Xi$, the
virtual normal bundle to $\Xi$ is given by the moving part of the above 
complex with respect to the torus action \cite{GP}. The fixed part determines 
the induced virtual cycle $[\Xi]^{vir}$. 
The local contribution of a fixed locus represented by a graph 
$\Gamma$ can be evaluated using normalization exact sequences \cite{GP,K}. 
In order to write down the answer in concise from, 
let us define a flag \cite{K} to be pair $(v,e)\in V(\Gamma)\times E(\Gamma)$
so that $v$ lies on $e$. 
One can think of a flag as an oriented edge. We also define the valence 
of a vertex $v$ to be the number of flags $(v,e)$. 
Recall that a vertex represents either a point of $\Sigma$ or a 
connected union of irreducible components mapping to one of the fixed 
points $P_1,P_2\in X$. If $v$ represents a point $p_v$ we have either 
$val(v)=1$, if $p_v$ is a smooth point or $val(v)=2$ if $p_v$ is an 
ordinary double point. We will denote the set of all such vertices 
by $V_{02}(\Gamma)$.
Note that in the last case, $p_v$ is the transverse intersection
point of two horizontal components $\Sigma_{e_1(v)}, \Sigma_{e_2(v)}$
corresponding to the flags $(v,e_1(v)), (v,e_2(v))$. 
In all other cases, $v$ represents a stable genus $g_v$ curve 
$\Sigma_v$ with $val(v)$ marked points
$(p_1,\ldots, p_{val(v)})$. The markings are points of intersection of 
$\Sigma_v$ with horizontal components $\Sigma_{e_k(v)}$
where $(v,e_k(v))$, $k=1,\ldots,val(v)$ are all flags containing $v$. 
The fixed locus $\Xi$ is isomorphic to a quotient of the direct product $\prod_{v\in V(\Gamma) \setminus
V_{02}(\Gamma)}\om_{g_v,val(v)}$ by a finite group $G$. $G$ is the automorphism group of an 
arbitrary fixed map in $\Xi$ and it fits in the exact sequence
\begin{equation}\label{eq:autgrp}
1\longrightarrow\prod_{e\in E(\Gamma)}\IZ/d_e\longrightarrow G\longrightarrow {\rm Aut}(\Gamma)\longrightarrow 1,\,
\end{equation}
where $Aut(\Gamma)$ is the automorphism group of the graph $\Gamma$. 
Note that each marked point $p_k$ determines a Mumford class 
$\psi_{k}\in H^*(\om_{g_v,val(v)})$. 

We have a normalization exact sequence of the form 
\be\label{eq:normA} 
\begin{split} 
& 0 \ra f^*T_X \ra \bigoplus_{e\in E(\Gamma)} f^*_eT_X \oplus \bigoplus_{v\in 
V(\Gamma)} f^*_vT_X \ra \bigoplus_{v\in V(\Gamma)} (T_{P_{k_v}}X)^{val(v)}
\ra 0.\cr
\end{split}
\ee
The associated long exact sequence reads 
\be\label{eq:normB}
\begin{split} 
0 & \ra H^0(\Sigma, f^*T_X) \ra 
\bigoplus_{e\in E(\Gamma)}H^0(\Sigma_e, f^*_eT_X)\oplus \bigoplus_{v\in 
V(\Gamma)} T_{P_{k_v}}X 
 \ra \bigoplus_{v\in V(\Gamma)} (T_{P_{k_v}}X)^{val(v)} \cr
& \ra 
H^1(\Sigma, f^*T_X)\ra \bigoplus_{e\in E(\Gamma)}H^1(\Sigma_e, f^*_eT_X)
\oplus \bigoplus_{v\in V(\Gamma)} H^1(\Sigma_v, \CO_{\Sigma_v})\otimes 
T_{P_{k_v}}X\ra 0.\cr
\end{split}
\ee
The moving part of the automorphism group $Aut(\Sigma)$ consists of 
holomorphic vector fields on the horizontal components $\Sigma_e$ 
which vanish at the points of $\Sigma_e$ which are nodes of $\Sigma$. 
let $\nu_e$ denote the divisor of such points on $\Sigma_e$. Then we have 
\be\label{eq:autA}  
Aut(\Sigma)^m = \oplus_{e\in E(\Gamma)} H^0(\Sigma_e, T_{\Sigma_{e}}(-\nu_e)).
\ee
The moving part of $Def(\Sigma)$ consists only of deformations 
of the nodes which lie at least on one horizontal component.
That is we have 
\be\label{eq:defA}
Def(\Sigma)^m= \bigoplus_{\substack{v\in V_{02}(\Gamma)\\ val(v)=2}} 
T_{p_v}(\Sigma_{e_1(v)})\otimes T_{p_v}(\Sigma_{e_2(v)})
\oplus \bigoplus_{v\in V(\Gamma)\setminus V_{02}(\Gamma)} 
\left(\oplus_{k=1}^{val(v)} T_{p_k}\Sigma_{e_k(v)} \otimes
T_{p_k}\Sigma_v\right).
\ee
Therefore the $T$-equivariant K-theory class of the virtual normal 
bundle to $\Xi$ is given by 
\begin{eqnarray}
[N_\Xi^{\rm vir}]&=& \sum_{e\in E(\Gamma)} \left(-[H^1(\Sigma_e,f^*T_X)^m]
+[H^0(\Sigma_e, f^*T_X)^m]-[H^0(\Sigma_e,T_{\Sigma_e})^m]\right)+
\nonumber\\&&
+\,\sum_{\substack{v\in V_{02}(\Gamma)\\ val(v) =2}}\left(
-[T_{P_{k_v}}X]+[T_{p_v}\Sigma_{e_1(v)}]+[T_{p_v}\Sigma_{e_2(v)}]
+[T_{p_v}\Sigma_{e_1(v)}\otimes T_{p_v}\Sigma_{e_2(v)}]\right)+
\nonumber\\&&
+\,\sum_{\substack{v\in V(\Gamma)\setminus V_{02}(\Gamma)\\}}
\left(-[H^1(\Sigma_v, f^*T_X)]+\sum_{k=1}^{val(v)}[T_{p_k}\Sigma_v\otimes 
T_{p_k}{\Sigma_{e_k(v)}}]+\sum_{k=1}^{val(v)}[T_{p_k}{\Sigma_{e_k(v)}}]\right)\nonumber\\&&
-\, \sum_{\substack{v\in V(\Gamma)\setminus V_{02}(\Gamma)\\}}\left((val(v)-1)[T_{P_{k_v}}X]\right).
\label{eq:locB}
\end{eqnarray}
The local contribution of $\Xi$ is of the form
\begin{equation}\label{eq:locC}
\int_{[\Xi]^{\rm vir}} \frac{1}{e_{T}(N^{\rm vir}_\Xi)} = \frac{1}{|{\rm Aut}(\Gamma)|}\prod_{e\in
E(\Gamma)}F(e)\prod_{v\in V_{02}(\Gamma)} G(v) \prod_{v\in
V(\Gamma)\setminus V_{02}(\Gamma)} \int_{(\om_{g_v,val(v)})_T} H(v)\,,
\end{equation}
where the expressions $F(e),G(v),H(v)$ are given as follows.  For any
vertex $v$ we denote by $\rho_{k_v}^{1,2,3}$ the weights of the torus
action on the holomorphic tangent space to $X$ at $P_{k_v}$.  Hence,
for $k_v=1$ we have $(\rho_1^1, \rho_1^2,\rho_1^3) =
(-\lb_z,-\lb_u,-\lb_v)$ and for $k_v=2$, $(\rho_2^1,
\rho^2_2,\rho_2^3) = (\lb_z,-\lb_u-\lb_z,-\lb_v-\lb_z)$.  We have
\begin{eqnarray}
F(e)&=&\frac{(-1)^{d_e-1}}{d_e(d_e!)^2}\prod_{k=1}^{d_e-1}
\left(k+\frac{d_e\lb_u}{ \lb_z}\right)\left(k+\frac{d_e\lb_v}{ \lb_z}\right)
\nonumber\\
G(v)&=&\left\{\begin{array}{ll} 1 & {\rm{if}}\ v\in V_{02}(\Gamma),\ 
val(v) =1 \\ 
\displaystyle\frac{d_{e_1}^2d_{e_2}^2{\rho_{k_v}^1}{\rho_{k_v}^2}{\rho_{k_v}^3}}{
(d_{e_1}+d_{e_2})(\rho_{k_v}^1)^3} & {\rm{if}}\ v\in V_{02}(\Gamma),\
val(v)=2 
\end{array}\right.
\nonumber\\
H(v)&=&\left({\rho_{k_v}^1}{\rho_{k_v}^2}{\rho_{k_v}^3}H^3\right)^{val(v)-1}\frac{c_{g_v}(\IE^{\vee}(\rho^1_{k_v}H))
c_{g_v}(\IE^{\vee}(\rho^2_{k_v}H))
c_{g_v}(\IE^{\vee}(\rho^3_{k_v}H))}{
\prod_{k=1}^{val(v)}\frac{\rho^1_{k_v}H}{
d_{e_k}}\left(-\psi_k +\frac{\rho^1_{k_v}H}{
d_{e_k}}\right)},
\label{eq:locD} 
\end{eqnarray}
where $\IE$ is the Hodge bundle on the moduli space $\om_{g_v,val(v)}$,
and $H\in H^*_T(pt)$ generates the equivariant cohomology of a point.

Our goal is to derive a similar formula for a fixed locus in the moduli 
space of equivariant holomorphic maps. 
Suppose we are given a fixed locus $\Xi_s$ in the moduli
space of symmetric holomorphic maps described by a symmetric graph 
$\Gamma_s$. Recall that $\Gamma_s$ is endowed with a pair of 
involutions $\tau_E:E(\Gamma_s)\ra E(\Gamma_s)$, 
$\tau_V:V(\Gamma_s)\ra V(\Gamma_s)$ which satisfy conditions $(I)-(III)$ 
above. We can regard $\tau_E, \tau_V$ as generators of cyclic groups 
of order two $\langle \tau_E\rangle$, $\langle \tau_V\rangle$ 
acting on $E(\Gamma)$ and respectively $V(\Gamma)$. 

According to our previous discussion, $\sigma:\Sigma \ra 
\Sigma$ maps each component $\Sigma_v\in f^{-1}(P_1)$ to the component 
$\Sigma_{\tau_V(v)}\in f^{-1}(P_2)$. 
This map is an antiholomorphic isomorphism between 
$\Sigma_v$ and $\Sigma_{\tau_V(v)}$. 
Any deformation of the triple $(\Sigma, f,\sigma)$ 
must preserve this relation, 
therefore the fixed locus $\Xi_s$ is isomorphic to a quotient by a finite group 
of the direct product $\prod_{<v>\in V(\Gamma_s)/\langle \tau_V\rangle}{\om_{v,val(v)}}$.
This means that we take a single factor $\om_{v,val(v)}$ for each 
orbit of the group $\langle \tau_V\rangle$ in $V(\Gamma_s)$. 

Next we have to determine the virtual normal bundle to $\Xi_s$. 
The main point is that we have to take into account only infinitesimal 
deformations of $(\Sigma,f,\sigma)$ which preserve the symmetric 
structure and the equivariance condition $f=I\circ f \circ \sigma$.  
This will impose constraints on the obstructions as well. 
An important observation is that $\sigma$ induces antiholomorphic involutions
on the cohomology groups $H^*(\Sigma, T_\Sigma)$. Similarly, the pair 
$(I,\sigma)$ induces antiholomorphic involutions on 
$H^*(\Sigma, f^*T_\Sigma)$. The tangent-obstruction complex of a 
symmetric map $(\Sigma, f,\sigma)$ can be obtained by taking 
the invariant part of \refeq{tangobsA}
\be\label{eq:tangobsB} 
\begin{aligned} 
0& \ra H^0(\Sigma, T_\Sigma)^\sigma\ra H^0(\Sigma, f^*T_X)^{(I,\sigma)} 
\ra {\BT}_s^1 
\ra H^1(\Sigma, T_\Sigma)^\sigma
 \ra H^1(\Sigma, f^*T_X)^{(I,\sigma)} \ra {\BT}_s^2\ra 0.\cr
\end{aligned}
\ee
In order to determine the virtual normal bundle, we have to take the moving 
part of \refeq{tangobsB}. This is equivalent to taking the invariant 
part of the moving part of \refeq{tangobsA}. Let us consider the action 
of the antiholomorphic involution on all cohomology groups involved in 
\refeq{normB}, \refeq{autA} and \refeq{defA}. Recall that $\sigma$ 
permutes the components of $\Sigma$ according to the rules $(I)-(III)$ 
below fig. 1. In particular, a horizontal component $\Sigma_e$ is mapped 
either to itself or to another horizontal component of $\Sigma_{\tau_E(e)}$
equal degree. A vertical component $\Sigma_v$ is mapped to another 
vertical component $\Sigma_{\tau_V(v)}$ of the same genus. 

If $\tau_E(e)\neq e$, $\sigma$ induces an 
antiholomorphic involution on 
$$H^0(\Sigma_e,T_{\Sigma_e}(-\nu_e))\oplus 
H^0(\Sigma_{\tau_E(e)},T_{\Sigma_{\tau_E(e)}}(-\nu_e))$$ 
and we are instructed to take the 
invariant real subspace. This can be (non-canonically) identified with 
$H^0(\Sigma_e,T_{\Sigma_e}(-\nu_e))$ as a complex vector space. 
Note that 
$H^0(\Sigma_e,T_{\Sigma_e}(-\nu_e))$ and 
$H^0(\Sigma_{\tau_E(e)},T_{\Sigma_{\tau_E(e)}}(-\nu_e))$
are isomorphic as T-equivariant complex vector spaces, hence  
$H^0(\Sigma_e,T_{\Sigma_e}(-\nu_e))$ depends only on the orbit 
$\langle e \rangle \in E(\Gamma)/\langle\tau_E\rangle$. 
However, there still is an ambiguity in this process related to the choice 
of an orientation on the moduli space. As noticed before, in standard 
Gromov-Witten theory, the moduli spaces come equipped with a canonical 
orientation induced by the complex structure. Here, the fixed 
subspace does not carry a canonical complex structure. Therefore 
we could equally well 
identify it to the complex conjugate ${H}^0(\Sigma_e(-\nu_e))^*$, 
and this gives rise to a sign ambiguity in the evaluation of the contribution 
of the fixed locus. This ambiguity cannot be resolved in the absence 
of a rigorous construction of the moduli space and the virtual cycle. 
Similar problems will be encountered throughout the rest of this analysis
as well. The only available solution is to fix a set of conventions 
and keep in mind that the resulting contributions may be off by a 
sign, which will be fixed later. 

If $\tau_E(e)=e$, there is an antiholomorphic involution 
$H^0(\Sigma_e,T_{\Sigma_e}(-\nu_e))
\ra H^0(\Sigma,T_{\Sigma_e}(-\nu_e))$, 
and we have to take the fixed subspace. The details 
of this computation are given in appendix A. 

The action of $\sigma$ on the deformations \refeq{defA} can be easily inferred 
from the action on $\Sigma$, which exchanges the nodes pairwise. 
For any two nodes $p,p'$ forming an orbit of $\sigma$ the invariant 
deformations can be noncanonically identified to the deformation of 
$p$ (up to a complex conjugation). 

Similar considerations apply to the cohomology groups 
$H^{0,1}(\Sigma_e, f^*_eT_X)$ and $H^{0,1}(\Sigma_v, f_v^*T_X)$. 
The former are pairwise identified if $\tau_E(e)\neq e$, by analogy 
with the automorphism groups. 
For $\tau_E(e)=e$, the invariant subspaces are computed again in appendix A. 
The groups $H^{0,1}(\Sigma_v, f_v^*T_X)$ are also pairwise identified.
Given an orbit $(v,\tau_V(v))$, we can choose the invariant subspace to be 
$H^{0,1}(\Sigma_v, f_v^*T_X)$ with $k_v=1$. 
Collecting the facts, the equivariant K-theory class of the virtual normal
bundle $N_{\Xi_s}^{vir}$ can be written as follows
\begin{eqnarray}
[N_{\Xi_s}^{\rm vir}]&=& \sum_{\substack{\langle e\rangle \in E(\Gamma)/\langle 
\tau_E\rangle\\ \tau_E(e)\neq e\\}} \left(-[H^1(\Sigma_e, f_e^*T_X)^m]+
[H^0(\Sigma_e, f_e^*T_X)^m]-
[H^0(\Sigma_e,T_{\Sigma_e})^m]\right)+
\nonumber\\&&
+\,\sum_{\substack{e\in E(\Gamma)\\ \tau_E(e)=e\\}}
\left(-[H^1(\Sigma_e, f_e^*T_X)^{\sigma,m}]+
[H^0(\Sigma_e, f_e^*T_X)^{\sigma,m}]-
[H^0(\Sigma_e,T_{\Sigma_e})^{\sigma,m}]\right)+
\nonumber\\&&
+\,\sum_{\substack{v\in V_{02}(\Gamma)\\ val(v) =2, k_v=1}}\left(
-[T_{P_{k_v}}X]+[T_{p_v}\Sigma_{e_1(v)}]+[T_{p_v}\Sigma_{e_2(v)}]
+[T_{p_v}\Sigma_{e_1(v)}\otimes T_{p_v}\Sigma_{e_2(v)}]\right)+
\nonumber\\&&
+\,\sum_{\substack{v\in V(\Gamma)\setminus V_{02}(\Gamma), k_v=1\\}}
\!\!\!\left(\!-[H^1(\Sigma_v, f^*T_X)]\!+\!\sum_{k=1}^{val(v)}\![T_{p_k}\Sigma_v\otimes 
T_{p_k}{\Sigma_{e_k(v)}}]\!+\!\sum_{k=1}^{val(v)}\![T_{p_k}{\Sigma_{e_k(v)}}]
\!\right)\nonumber\\&&
-\,\sum_{\substack{v\in V(\Gamma)\setminus V_{02}(\Gamma), k_v=1\\}}\left((val(v)-1)[T_{P_{k_v}}X]\right).
\label{eq:locE}
\end{eqnarray}
The automorphism group $G_s$ is an extension given by the exact sequence
\begin{equation}\label{eq:autgrpii}
1\longrightarrow\prod_{e\in E(\Gamma)/\langle\tau_E\rangle}\IZ/d_e
\longrightarrow G_s\longrightarrow {\rm Aut}(\Gamma_s)\longrightarrow 1,\,
\end{equation}
where $Aut(\Gamma_s)$ is the group of automorphisms which preserve the involution.

The local contribution of the symmetric fixed locus $\Xi_s$ reads
\begin{equation}
\int_{[\Xi_s]^{\rm vir}} \frac{1}{ e_{T}(N^{\rm vir}_{\Xi_s})} =
\!\frac{1}{|{\rm Aut}(\Gamma_s)|}\prod_{\substack{\langle e\rangle \in E(\Gamma)/\langle \tau_E\rangle\\
\tau_E(e)\neq e\\}}F(e) \!\prod_{\substack{e\in E(\Gamma)\\
\tau_E(e)=e\\}} C(e) \!\prod_{\substack{v\in V_{02}(\Gamma)\\k_v=1}}
G(v)\! \prod_{\substack{v\in V(\Gamma)\setminus V_{02}(\Gamma)\\k_v=1}}
\int_{(\om_{g_v,val(v)})_T} H(v)\,,
\label{eq:locF} 
\end{equation}
where $F(e), G(v), H(v)$ are given in \refeq{locD}. The function
$C(e)$ is derived in appendix A
\begin{equation}\label{eq:locG}
C(e)= \frac{1}{ d_e(d_e!)} \prod_{k=1}^{d_e-1}\left(k+\frac{d_e\lb_v}{\lb_z}
\right).
\end{equation}

\subsection{Computations} 

Let us now perform some concrete computations and run a comparison 
test with large $N$ duality predictions. 
First note that on general grounds, the unoriented topological 
free energy for the model considered here 
has an expansion of the form 
\be\label{eq:freenA} 
\CF_{(X,I)}= \sum_{h\geq 0}\sum_{c\geq 1} g_s^{-\chi} \sum_{d\geq 1} 
C_{\chi,d}q^{d/2}.
\ee
where $\chi$ denotes the Euler characteristic of the unoriented 
Riemann surface 
$\Sigma/\langle \sigma\rangle$. 
We have $-\chi=2h-2+c$ where $h$ is the number of handles and $c$ is the 
number of crosscaps. If the covering surface $\Sigma$ has genus $g$, 
we have $2h-2+c=g-1$. Moreover, $d\geq 1$ denotes the degree of the 
map ${\widetilde f}:\Sigma/\langle \sigma\rangle\ra X/I$. 
Note that the coefficients $C_{\chi,d}$ depend only on $\chi$ and 
not $(h,c)$ taken separately. This is commonly referred to in the 
physics literature as trading a certain number of crosscaps for 
a certain number of handles. In fact we will show below that the
fixed loci in the moduli space of symmetric maps of genus $g$ 
may have different values of $(h,c)$ as long as the combination 
$2h-2+c$ stays the same.

Now, recall that large 
$N$ duality \cite{SV} predicts an exact formula for all genus unoriented 
topological amplitudes of the form 
\be\label{eq:predA}
\CF_{(X,I)}= \sum_{\substack{d\geq 1\\ d\ \rm{odd}}}
{q^{d/2}\over 2n{\rm sin}{dg_s\over 2}}.
\ee
Using the generating functional for modified Bernoulli numbers 
\be\label{eq:bernoulli} 
{t/2\over \sin(t/2)} = \sum_{k=0}^\infty b_k t^{2k},
\ee
we can rewrite \refeq{predA} as 
\be\label{eq:predB}
\CF_{(X,I)}= \sum_{\substack{d\geq 1\\ d\ \rm{odd}}}\sum_{h\geq 0} 
\sum_{\substack{c\geq 1\\ c\ \rm{odd}}} g_s^{2h-2+c}d^{2h-3+c}
b_{h+{c-1\over 2}}q^{d/2}.
\ee
This allows us to identify all the coefficients $C_{\chi,d}$ in 
\refeq{freenA}. Our goal is to reproduce these results using 
the ${\bf A}$-model approach developed so far. We start with $\IR\IP^2$ 
instantons, that is $\chi=-1$ and consider several values of 
$d$. 

$i)\ {(\chi,d)=(-1,1)}.$ There is a single fixed locus represented by a 
graph with one edge of degree $d_e=1$. This can be trivially evaluated 
to $C_{-1,1}=1$, which is the correct result. 

$ii)\ {(\chi,d)=(-1,3)}.$ There are two fixed loci represented by the 
symmetric graphs in fig. 2. The graphs in the first row are symmetric 
graphs as discussed above. In case $(a)$ we have an involution mapping 
the horizontal component onto itself. In case $(b)$ the involution exchanges  
the outer horizontal components and maps the middle component onto itself. 
This information can be conveniently encoded in the quotient graphs 
represented on the second row. Horizontal components with an {\tt x} attached to a 
vertex represent odd multicovers of $\IR\IP^2\ra \IR\IP^2$. 
The other horizontal
components correspond to even multicovers of $\IP^1\ra \IR\IP^2$.
The local contributions are 
\be\label{rptwoA}
\begin{split} 
& C_{-1,3}^{(a)}= 
{1\over 18}\left(1+{3\lambda_v\over \lambda_z}\right)\left(2+{3\lambda_v\over 
\lambda_z}\right),\qquad
C_{-1,3}^{(b)}= {1\over 2}{\lambda_u\lambda_v\over \lambda_z^2}.\cr
\end{split}
\ee

\begin{figure}[ht]
   \centering
   \scalebox{0.7}{\includegraphics[angle=0]{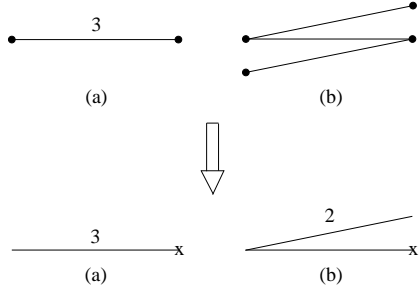}}
    \caption{Degree 3 symmetric graphs.}\label{degthree}
\end{figure} 

\noindent Taking into account the relation
\be\label{eq:toractCi}
\lb_z+\lb_u+\lb_v=0,
\ee 
it is straightforward to check that 
\be\label{eq:rptwoB}
C_{-1,3}=C_{-1,3}^{(a)}+C_{-1,3}^{(b)}={1\over 9},
\ee 
which matches the large $N$ duality result. Note that this is true for
any torus action compatible with the antiholomorphic involution.
As opposed to open string localization computations, the result is independent 
of the toric weights, which reflects the compactness of the moduli space.

$iii)\ {(\chi,d)=(-1,5)}.$ Here we find five symmetric graphs of total 
degree five represented (together with their quotient graphs) 
in fig. 3. 
\begin{figure}[ht]
    \centering
    \scalebox{0.7}{\includegraphics[angle=0]{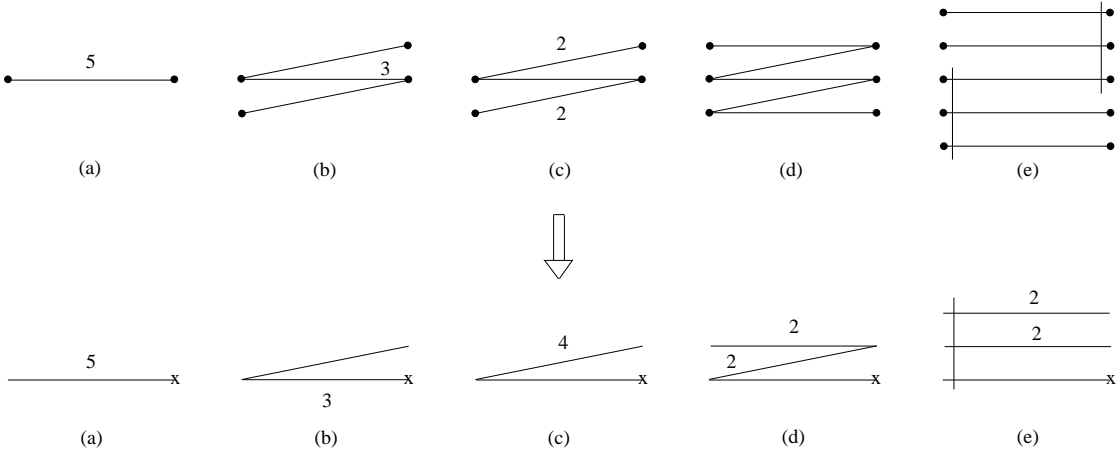}}
     \caption{Degree 5 symmetric graphs.}\label{degfive}
 \end{figure}

\noindent The local contributions are
\be\label{eq:rptwoC}
\begin{split}
&C_{-1,5}^{(a)} = {1\over 600}{24\lb_u^4-154\lb_u^3\lb_v+269\lb_v^2\lb_u^2
-154\lb_u\lb_v^3+24\lb_v^4\over \lb_z^4}, \qquad C_{-1,5}^{(c)} = {1\over 6}{\lb_u\lb_v(\lb_u-\lb_v)^2\over \lb_z^4},\\ 
&C_{-1,5}^{(b)} = {1\over 8}{\lb_u\lb_v(2\lb_u^2-5\lb_u\lb_v+2\lb_v^2)
\over \lb_z^4}, \qquad  C_{-1,5}^{(d)} = {1\over 4} {\lb_u^2\lb_v^2\over \lb_z^4}, \qquad
C_{-1,5}^{(e)} = {1\over 2} {\lb_u^2\lb_v^2\over \lb_z^4}.\\\
\end{split}
\ee
Using the condition \refeq{toractCi}, we obtain 
\be\label{rptwoD}
C_{-1,5}= C_{-1,5}^{(a)}+\cdots +C_{-1,5}^{(e)} = {1\over 25}.
\ee
This result is again independent of the choice of a torus action 
which preserves the involution. 

The higher degree computations become more cumbersome because 
the number of graphs increases rapidly. For example at degree seven, 
there are thirteen symmetric graphs which yield 
$C_{-1,7}={1\over 49}$ as expected. Although we do not have a general proof 
for any degree (at least for arbitrary values of torus weights subject 
to \refeq{toractA}), this is strong supporting evidence for our approach. 
The computation simplifies dramatically by making a special choice
of torus weights. However, before discussing this option, 
it may be instructive to 
perform higher genus computations for arbitrary weights. 
Note that if $(\chi,d)=(0,1)$, $C_{\chi,d}$ is trivially zero. 

$iv)\ (\chi,d)=(0,2)$. This is an interesting case. Topologically, 
we have a single symmetric graph with two edges of degree one and 
two vertices as shown in fig. 4. 
\begin{figure}[ht]
    \centering
    \scalebox{0.7}{\includegraphics[angle=0]{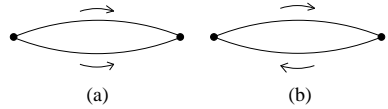}}
     \caption{$(\chi,d)=(0,2)$ symmetric graphs.}\label{degztwo}
 \end{figure}

\noindent However, we should remember at this point that the moduli space 
classifies triples $(\Sigma,f,\sigma)$ up to equivalence. In the present 
case, $\Sigma$ is supposed to be a prestable curve of genus $g=1$, 
and $\sigma:\Sigma \ra \Sigma$ a freely acting antiholomorphic involution. 
In order to obtain a fixed map under the torus action, $\Sigma$ 
must degenerate to a nodal curve of arithmetic genus one as in fig. 4. 
We claim that this curve admits two inequivalent antiholomorphic involutions. 
Let $w_1, w_2$ denote affine coordinates on the two components such that 
the restrictions of $f$ to the two components are $z=w_1$, $z=w_2$. 
One can define the following antiholomorphic involutions 
\be\label{eq:ahinvolD}
\begin{split}
& \sigma_1:\quad  w_1\ra -{1\over \bw_1},\quad w_2\ra -{1\over \bw_2}\cr
& \sigma_2:\quad  w_1\ra -{1\over \bw_2},\quad w_2\ra -{1\over \bw_1}.\cr
\end{split}
\ee
The domain $\Sigma$ has an automorphism $\phi:\Sigma \ra \Sigma$ 
which exchanges the two 
components. One can check that $\phi$ preserves both involutions 
$\sigma_1, \sigma_2$, hence the triples $(\Sigma,f,\sigma_1)$, 
$(\Sigma,f,\sigma_2)$ are not equivalent. They represent distinct fixed 
points of the moduli space, each having an automorphism group of order two. 
Both local contributions are equal to ${\lb_u\lb_v\over 4\lb_z^2}$ 
up to a sign ambiguity, which has been discussed in the previous section. 
We do not know how to fix this ambiguity from the first principles, 
so we have to rely on duality predictions and the compactness assumption. 
Large $N$ duality predicts a zero result for this amplitude. Moreover, 
by compactness of the moduli space, the result has to be 
independent of toric weights. This can be achieved only if 
the two contributions have a relative minus sign, in which case the 
result is trivial. We will simply adopt this rule as part of our 
computational definition of the virtual cycle. Additional 
supporting evidence will be found below.   

$v)\ (\chi,d)=(1,1)$. We have $2h-2+c=1$, and $c=1$, since the degree is 
one. This implies that $h=1$, and there is a single graph represented below.
\begin{figure}[ht]
    \centering
    \scalebox{0.7}{\includegraphics[angle=0]{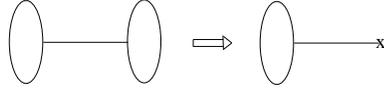}}
     \caption{$(\chi,d)=(1,1)$ symmetric graphs.}\label{degoneone}
 \end{figure}
Note that we have represented the genus one vertex by an ellipse 
in order to facilitate an intuitive understanding of the invariant map. 
Similarly, genus zero vertices will be represented by a vertical line 
segment in the following. 
The corresponding invariant is  
\be\label{eq:oneoneA} 
\begin{split} 
C_{1,1}& ={1\over (-\lb_zH)}\int_{\om_{1,1}}
{c_1(\IE^{\vee}(-\lb_zH))
c_1(\IE^{\vee}(-\lb_uH))
c_1(\IE^{\vee}(-\lb_vH))\over
\left(-\psi_1 -{\lb_z H}\right)}\cr
& = {\lb_u\lb_v\over \lb_z^2} \int_{\om_{1,1}} \psi -
{\lb_z\lb_u+\lb_u\lb_v+\lb_v\lb_z\over \lb_z^2} 
\int_{\om_{1,1}}\lambda \cr
& = b_1.\cr
\end{split}
\ee 
Note that at the last step we had to use again the condition \refeq{toractCi}. This is the expected result. 

$vi)\ (\chi,d)=(1,3)$. This is a more interesting case. We have 
$2h-2+c=1$ and $d=3$ which can be realized either as $(h,c)=(1,1)$ 
or as $(h,c)=(0,3)$. Therefore we can have surfaces with one handle 
and one crosscap or surfaces with no handles, but three crosscaps, 
as shown in fig. 6. 
\begin{figure}[ht]
    \centering
    \scalebox{0.9}{\includegraphics[angle=0]{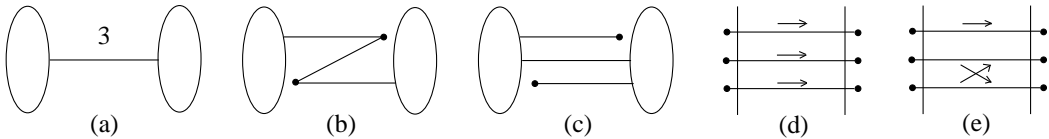}}
     \caption{$(\chi,d)=(1,3)$ symmetric graphs.}\label{}
 \end{figure}
The local contributions of these fixed loci read 
\be\label{eq:onethreeA}
\begin{aligned}[b]
& C_{1,3}^{(a)}=-{b_1\over 2}
{5\lb_z^2\lb_u\lb_v-2\lb_z^4+18\lb_v^2\lb_u^2\over
\lb_z^4},& &C_{1,3}^{(b)}={b_1\over 2}{\lb_u\lb_v\over \lb_z^2},& \\ 
&C_{1,3}^{(c)}={b_1}{\lb_u\lb_v(\lb_u\lb_v-2\lb_u\lb_z-2\lb_v\lb_z)\over 
\lb_z^4},& &C_{1,3}^{(d)}=-{1\over 6}{\lb_u^2\lb_v^2\over \lb_z^4},&\\
&C_{1,3}^{(e)}={1\over 2}{\lb_u^2\lb_v^2\over \lb_z^4}.& & &
\end{aligned}
\ee
Some comments are in order here. In the process of evaluating 
graphs $(a)$, $(b)$ and $(c)$ one has to integrate polynomials in  
tautological and Hodge classes on $\om_{1,1}$ and $\om_{1,2}$. 
This can be done using string and dilaton equations as explained 
for example in \cite{EW}. We omit the details since this is standard 
material. Graphs $(d)$ and $(e)$ are more subtle. It is clear that the
domain $\Sigma$ is the same in both cases. However, we obtain again 
two distinct fixed points because we can define inequivalent involutions. 
Let $w_1, w_2, w_3$ denote local affine coordinates on the three 
horizontal components so that the map $f$ is given by $z=w_1,\ z=w_2,\
z=w_3$. Note that the automorphism group of $\Sigma$ is isomoprhic 
to $S_3$ (the permutation group of three letters). 
First, we have an antiholomorphic involution
\be\label{eq:ahinvolE} 
\sigma_1:\quad w_1\ra -{1\over \bw_1},\quad  w_2\ra -{1\over \bw_2},\quad 
 w_3\ra -{1\over \bw_3}
\ee 
which is preserved by the whole automoprhism group $S_3$. 
However, one can define three more antiholomorphic involutions 
\be\label{eq:ahinvolF}
\begin{split}
& \sigma_2:\quad  w_1\ra -{1\over \bw_1},\quad w_2\ra -{1\over \bw_3},\quad 
 w_3\ra -{1\over \bw_2}\cr
& \sigma_3:\quad  w_1\ra -{1\over \bw_3},\quad w_2\ra -{1\over \bw_2},\quad 
 w_3\ra -{1\over \bw_1}\cr
& \sigma_4:\quad  w_1\ra -{1\over \bw_2},\quad w_2\ra -{1\over \bw_1},\quad 
 w_3\ra -{1\over \bw_3}.\cr
\end{split}
\ee
Writing $S_3$ as $\IZ_2\times \IZ_3$, 
one can check that each such involution is preserved by the $\IZ_2$ subgroup, 
and $(\sigma_2, \sigma_3,\sigma_4)$ form an orbit of the $\IZ_3$ subgroup. 
Therefore the triples $(\Sigma,f,\sigma_{2,3,4})$ are equivalent,
and represent 
the same fixed point in the moduli space. This point has an automorphism 
group of order two. Finally, the contributions of these two graphs must be 
assigned different signs, as shown in equation \refeq{onethreeA}. 
This assignement is consistent with the sign choices made in the 
previous example. In fact, we can infer a simple rule for fixing the signs 
of graphs of this type. The difference between $\sigma_1$ and $\sigma_{2,3,4}$ 
is that $\sigma_1$ does not permute the horizontal components, while 
$\sigma_{2,3,4}$ involve a transposition $\tau$. The sign is simply 
given by $-$(signature of $\tau$). All sign rules obtained so far can be neatly summarized in the 
formula $(-1)^{\left[\frac{c-1}{2}\right]}$, where $[r]$ denotes 
the greatest integer smaller or equal to a given rational number $r$. We will show below that this 
rule also works for $(\chi,d)=(0,4),(1,5)$, therefore it is tempting 
to conjecture its validity for arbitrary $\chi,d$. 

To conclude this example, note that the sum of all contributions listed in 
\refeq{onethreeA} is $C_{1,3}=b_1={1\over 24}$ independent of the toric 
weights. This is again in agreement with large $N$ duality. 

$vii)\ (\chi,d)=(0,4)$. This case is very similar to $(\chi,d)=(0,2)$. 
We have five graphs represented in fig. 7. 
\begin{figure}[ht]
    \centering
    \scalebox{1.1}{\includegraphics[angle=0]{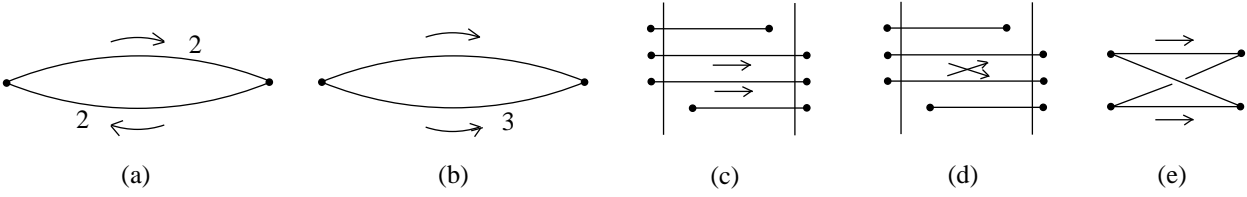}}
     \caption{$(\chi,d)=(0,4)$ symmetric graphs.}\label{degzfour}
 \end{figure}

\noindent The individual contributions are 
\be\label{eq:zerofourA} 
\begin{aligned} 
& C_{0,4}^{(a)}= -{1\over 4} {\lb_u\lb_v(\lb_z^2-4\lb_u\lb_v)\over \lb_z^4},& 
& C_{0,4}^{(b)}={1\over 8}{\lb_u\lb_v(2\lb_z^2-9\lb_u\lb_v)\over \lb_z^4},& & &\cr
& C_{0,4}^{(c)}={1\over 2}{\lb_u^2\lb_v^2\over \lb_z^4},& 
& C_{0,4}^{(d)}=-{1\over 2}{\lb_u^2\lb_v^2\over \lb_z^4},&
& C_{0,4}^{(e)}={1\over 8}{\lb_u^2\lb_v^2\over \lb_z^4}.&\cr
\end{aligned}
\ee
Note that for the first graph, the only admissible involution has to 
permute the two components in order to obtain a symmetric map. 
We have also used the above set of rules for fixing the signs 
of graphs three and four. The sum of all individual contributions 
is zero, as predicted by large $N$ duality. 
Finally, let us consider our last example 

$ix)\ (\chi,d)=(1,5)$. This computation is more involved because 
it involves alltogether fifteen symmetric graphs, which are 
represented in fig. 8. 
\begin{figure}[ht]
    \centering
    \scalebox{0.9}{\includegraphics[angle=0]{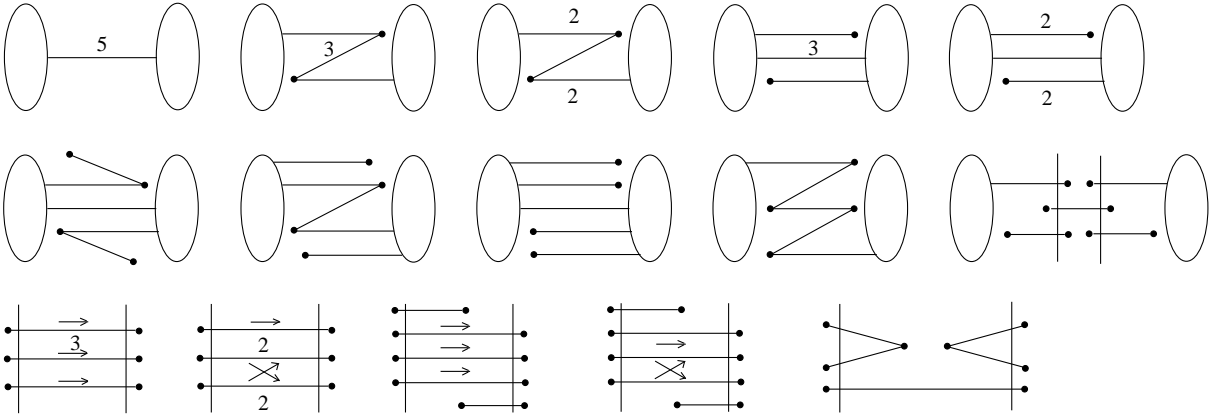}}
     \caption{$(\chi,d)=(1,5)$ symmetric graphs.}\label{degonef}
 \end{figure}
Since the formulae become quite cumbersome, we will not list all 
individual contributions here. Let us just note that using the 
rules specified so far one can obtain the final answer to be 
$C_{1,5}=b_1$, again in precise agreement with duality predictions. 
Given the complexity of the cancellations involved in the process, 
this is a highly nontrivial test of our approach. 

We conclude this section with a last remark on the 
free energy of this model. Recall that the duality prediction for 
this quantity is \refeq{freenA}. As briefly mentioned earlier, one 
can reproduce this formula by localization techniques. The main point, 
as shown in \cite{FP,YM} is to make a good choice of weights so that a large 
number of graphs vanish, and the remaining graphs can be easily evaluated. 
In our case, we can choose for example $\lb_u=\lb_v+\lb_z=0$. A consequence 
of this choice is that only graphs with exactly one edge have 
nonzero contributions. This phenomenon has been noticed in the 
context of closed strings in \cite{FP,YM} and in the context of 
open strings in \cite{KL,LS}. Then one can easily recover the formula 
\refeq{freenA} using essentially the same manipulations as in 
the above references. We will not reproduce the details here. 

\section{Localization and Invariant Graphs II} 

So far we have proposed a concrete approach to unoriented 
closed string enumerative geometry, and applied it in a simple 
context -- the resolved conifold geometry. We claim that this 
formalism is in fact applicable to any toric Calabi-Yau 
threefold with antiholomorphic involution. In order to illustrate 
the method, in this section we will perform localization 
computations for a more complicated local geometry. The $\IR\IP^2$ 
results will be succesfully checked against mirror symmetry computations. 

The threefold $X$ is now described by the following toric data 
\begin{equation}\label{eq:toricB} 
\begin{array}{ccccccc}
& X_1 & X_2 & X_3 & X_4 & X_5 & X_6\cr
\IC^* & 1 & -3 & 1 & 1 & 0 & 0\cr
\IC^* & 0 & 1 & -1 & -1 & 1 & 0 \cr
\IC^* & 0 & 0  & 1 & 1 & -3 & 1.\cr
\end{array}
\end{equation}

\begin{figure}[ht]
\centerline{\epsfxsize=11.5cm \epsfbox{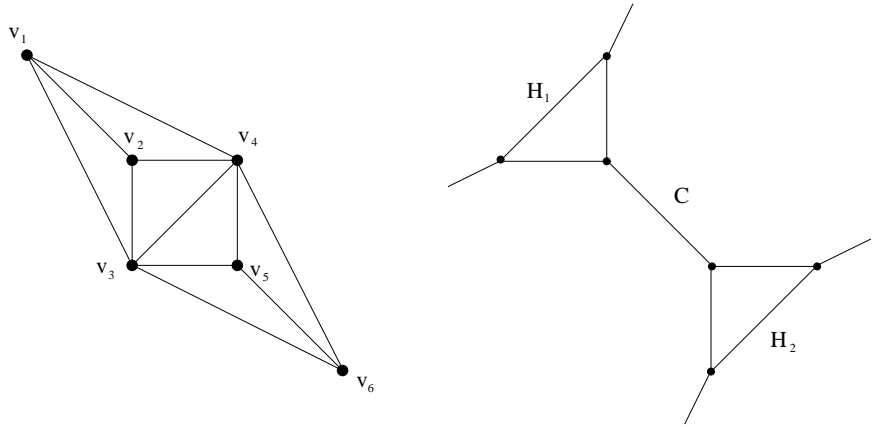}}
\caption{Toric diagram and configuration of invariant curves for the 
model given in \refeq{toricB}.} 
\label{fig:2P2sandaP1} 
\end{figure}

\noindent
Note that there are two compact divisors $D_1, D_2$ on $X$, 
each isomorphic to $\IP^2$, and a rational $(-1,-1)$ curve $C$
which intersects both divisors transversely. The divisors do not 
intersect each other.  
 
The antiholomorphic involution is given by 
\be\label{eq:ahinvolFi}
I:(X_1,X_2,X_3,X_4,X_5,X_6) \ra (\BX_6,\BX_5,\BX_4,-\BX_3,-\BX_2,\BX_1).
\ee
We follow the steps described in the previous section. Consider a 
real torus action on $X$ of the form 
\be\label{eq:toractDi} 
e^{i\phi}\cdot(X_1,X_2,X_3,X_4,X_5,X_6)\ra(e^{i\lb_1\phi}X_1, 
e^{i\lb_2\phi}X_2, e^{i\lb_3\phi}X_3, e^{i\lb_4\phi}X_4, e^{i\lb_5\phi}X_5,
e^{i\lb_6\phi}X_6).
\ee
This torus action is compatible with the antiholomorphic involution 
if the following relations between weights are satisfied 
\be\label{eq:toractE}
\lb_1+\lb_6=0,\quad \lb_2+\lb_5=0,\quad \lb_3+\lb_4=0.
\ee
The configuration of $T$-invariant curves on $X$ is represented in fig. 9. 
We have the familiar configurations of invariant hyperplanes in each 
$\IP^2$ connected by the curve $C$. 

The second homology of $X$ is generated by the class of $C$ and the two 
hyperplane classes of $D_1, D_2$, which will be denoted by $H_1, H_2$.
Note that $I$ maps $D_1$ antiholomorphically onto $D_2$, and 
it acts as $I_*(C) = -C,\ I_*(H_1)=-H_2$ on  
homology. Therefore the second homology 
of the quotient space $X/I$ is generated by $C/I$, which is an 
$\IR\IP^2$ cycle as before, and a second class $H$. 
The K\"ahler class $\omega$ is compatible with the action on homology 
if $I^*\omega= -\omega$.

The closed string Gromov-Witten invariants of $X$ can be computed 
by localization as explained before. The algorithm is very similar to the 
previous case, except that the structure of the fixed loci is more 
complicated. In fact one can extend without too much effort the graph 
representation to the present case \cite{K}. The same labeling rules 
apply except that the integer $k_v$ can now take values $k_v=1,\ldots, 6$ 
corresponding to the six fixed points $P_1, \ldots, P_6$ of the torus action 
on $X$. The edges $e\in E(\Gamma)$ represent irreducible rational components 
of $\Sigma$ which are mapped $d_e:1$ to one of the invariant  curves on $X$.
For convenience, an edge $e$ will be called of type $(ij)$, with 
$i,j=1,\ldots,6$ is the image $f(\Sigma_e)$ in $X$ is the invariant curve 
${\overline {P_iP_j}}$. We will also denote by $E_{(ij)}(\Gamma)$ 
the set of edges of $\Gamma$ of given type $(ij)$.
The local contribution of an arbitrary graph can be worked as in the 
previous section using normalization exact sequences. We will not attempt 
to write down an explicit formula since it would be too complicated. 
Concrete computations will be performed in several examples below. 

In the unoriented sector of the theory, we have to sum over triples 
$(\Sigma, f,\sigma)$ as in above. This means that the corresponding fixed 
loci can be classified by symmetric graphs.  A symmetric graph is defined 
to be a Kontsevich graph equipped with two involutions $\tau_E:E(\Gamma) 
\ra E(\Gamma)$, $\tau_V:V(\Gamma)\ra V(\Gamma)$ subject to certain 
conditions analogous to $(I)-(III)$ in the previous section. 
Let us make this more explicit. 

$(I')$ The action of $\tau_V$ on vertices should be compatible with the target 
space involution $I$, $k_{\tau_V(v)} = k_v+3$, and it should 
leave the genus invariant, $g_{\tau_V(v)}=g_v$.

$(II')$ The action of $\tau_E$ on edges should be compatible with 
target space involution. This means that there should be a precise 
correlation between the type of $e$ and the type of $\tau_E(e)$ 
as follows
\be\label{eq:correlA} 
\begin{split} 
& e\in E_{(14)}(\Gamma) \Rightarrow \tau_E(e) \in E_{(14)}(\Gamma)\cr
& e\in E_{(12)}(\Gamma) \Rightarrow \tau_E(e) \in E_{(45)}(\Gamma)\cr
&  e\in E_{(13)}(\Gamma) \Rightarrow \tau_E(e) \in E_{(46)}(\Gamma)\cr
&  e\in E_{(23)}(\Gamma) \Rightarrow \tau_E(e) \in E_{(56)}(\Gamma).\cr
\end{split} 
\ee
Moreover, $\tau_E$ should leave the degree invariant $d_{\tau_E(e)}=d_e$
for any $e\in E(\Gamma)$; if $\tau_E(e)=e$, $d_e$ should be odd. 

$(III')$ The involutions $\tau_V$, $\tau_E$ must be compatible i.e. if 
$(v,e)$ is a flag, then $(\tau_V(v), \tau_E(e))$ must be also a flag. 

Given these conditions, it is fairly straightforward to work out the 
local contribution of a symmetric graph. Essentially, the new element 
compared to the previous section is the presence of extra vertices and 
edges mapping to $D_1, D_2$, but not $C$. A vertex or edge mapping to 
$D_1$ is always pairwise identified to a vertex or edge mapping 
to $D_2$. For each such pair of vertices $(v,\tau_V(v))$ we will 
include a single integral $\int_{\om_{g_v,val(v)}}$ by analogy with 
the previous case. Similarly for each pair of edges 
$(e,\tau_E(e))\in \left(E(\Gamma) \setminus E_{(14)}(\Gamma)\right)^2$  
we include only one edge factor. If the torus action is compatible 
with the involution $I$, the edge factors of $e$ and $\tau_E(e)$ 
are equal, hence there is no ambiguity here. 
If $e\in E_{(14)}(\Gamma)$, 
then $f(\Sigma_e)=C$, and we apply the rules of the previous 
section. Therefore the only ambiguities present in this 
approach are the ones encountered in the previous computation
for which have a simple set of rules. We will show in the following that
the same set of rules applies to the present case without modification. 
This is a nontrivial consistency check of the methods developed here. 

\subsection{Results} 

Let us now present some results obtained using this algorithm. 
The unoriented free energy has an expansion of the form 
\be\label{eq:freenB} 
\CF_{(X,I)} = \sum_{h\geq 0} \sum_{c\geq 1} g_s^{-\chi}
\sum_{\substack {m,n\geq 0\\ (m,n)\neq 0\\}} C_{\chi,m,n} 
q_1^{n}q_2^{m/2}
\ee
where $(m,n)$ are defined by $f_*[\Sigma]=mC+n(H_1+H_2)$. 
Note that any symmetric map must have equal degrees with respect 
to $H_1, H_2$. The coefficients $C_{\chi,m,n}$ can be computed 
by summing over all symmetric graphs compatible with the triple 
$(\chi,m,n)$. This is a fairly straightforward, although tedious 
process. If we carefully employ all the rules found in the 
previous section,  we find the following expansion 
\be\label{eq:freenC}
\begin{split} 
\CF_{(X,I)} = & \frac{1}{g_s}\bigg(q_2^{1/2} -2q_1q_2^{1/2} + 5 q_1^2q_2^{1/2} 
-32q_1^3 q_2^{1/2} +  \ldots +{1\over 9}q_2^{3/2}-3q_1^2q_2^{3/2} +{268\over 9}q_1^3q_2^{3/2}
+\ldots \bigg)\cr
& +g_s\bigg({1\over 24} q_2^{1/2} -{1\over 12} q_1q_2^{1/2}  +\ldots \bigg).
\cr
\end{split}
\ee 
A couple of remarks are in order here. Note that all coefficients 
$C_{\chi,m,n}$ are independent of weights and satisfy the expected 
integrality properties. This is a nontrivial result, given the fact that 
the number of graphs increases very rapidly, and the local contributions 
become quite complicated. For example the coefficient $C_{-1,3,3}={268\over
9}$, 
which has the correct multicover behavior, 
is the result of summing one hundred and seventy-eight symmetric graphs, 
which can be naturally grouped in pairs. 
A typical pair of graphs that occurs in this computation is represented below. 
The local contribution of this pair is of the form 

\begin{figure}[ht]
\centerline{\epsfxsize=11cm \epsfbox{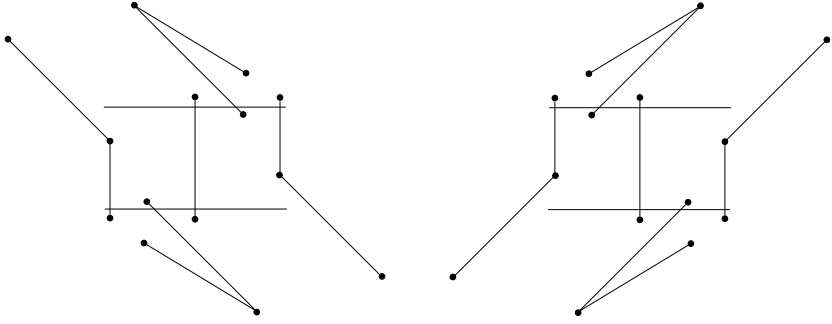}}
\caption{A pair of $(\chi,m,n)=(-1,3,3)$ symmetric graphs.} 
\label{fig:somegraphs} 
\end{figure}

\be\label{eq:monethreethreeA} 
{1\over 2}{-3\lb_u^5\lb_v+2\lb_u^4\lb_v^2+\lb_u^6-3\lb_u\lb_v^5+
2\lb_u^2\lb_v^4+\lb_v^6\over \lb_z^2\lb_u^2\lb_v^2}.
\ee 

Next, several coefficients on this expansion turn out to be zero, 
as a result of nontrivial cancellations among different graphs.
For example we have $C_{-1,3,1}=C_{-1,5,1} =0$. 
At higher genus, $\chi=1$ we obtain the expected multicover 
formulae, that is $C_{1,1,1}=-{1\over 12}=C_{-1,1,1}b_1$, and 
$C_{1,3,1}=0=C_{-1,3,1}b_1$. In the last case one has to sum over 
fifteen pairs of graphs, many of them involving integrals over 
$\om_{1,k}$ with various $k$. Moreover, at this level, we have 
also to use the fix signs as discussed in the previous section. 
In principle, one could compute higher order terms in the expansion,
but the computations become very tedious. We will not further 
pursue this aspect here. Below we compare the above results with $\chi=-1$ 
to mirror symmetry computations. 

\subsection{Mirror Symmetry} 

In this section we perform a {\bf B}-model computation for the
$\IR\IP^2$ free energy of the above model. We will find that 
the {\bf B}-model expansion is in precise agreement with the {\bf A}-model 
computation 
\refeq{freenC} up to an overall factor of $2$ which has been explained 
in section two (second paragraph below equation (2.4).)

The local mirror geometry of $X$ is a Landau-Ginzburg 
model described in terms of six dual 
variables $y_1,\ldots,y_6\in (\IC^*)^6$ subject to the 
following relations \cite{CKYZ,HV} 
\be\label{eq:relMori}
\begin{gathered}
y_1y_3y_4=e^{-t_1}y_2^3,\\
y_2y_5=e^{-t_2}y_3y_4,\\
y_3y_4y_6=e^{-t_3}y_5^3.
\end{gathered}
\ee
Here $t_1,t_2,t_3$ are complexified K\"ahler parameters, and the relations 
follow from the generators of the Mori cone of $X$. 
The Landau-Ginzburg superpotential is 
\be\label{eq:LGsup}
W_{LG}=\sum_{i=1}^6y_i.
\ee
We work in the patch $y_4=1$ and set $y_2=e^{v},y_3=e^{u}$, 
with $u,v\sim u+2\pi,v+2\pi$. Using the relations \eqref{eq:relMori}, we 
can now rewrite the local mirror equation \eqref{eq:LGsup} as
\be\label{eq:lmir}
e^{-t_1+3v-u}+e^{-t_2+u-v}+e^{-3t_2-3t_3+2u-3v}+e^u+e^v+1=0.
\ee
This Riemann surface determines a three dimensional hypersurface 
\be\label{eq:lmirt}
xz=e^{-t_1+3v-u}+e^{-t_2+u-v}+e^{-3t_2-3t_3+2u-3v}+e^u+e^v+1
\ee
in $(\BC^*)^2\times\IC^2$, which is the local mirror 
of $X$. 

Following the general considerations of \cite{AAHV}, 
the antiholomorphic involution $I:X\ra X$ should be mapped by 
mirror symmetry to a holomorphic involution $J$ of the 
hypersurface \refeq{lmirt}. Moreover, the $\IR\IP^2$ 
partition function (or superpotential) 
in the {\bf B}-model is related to a one-chain period 
of the holomorhic one-form $udv$ on the Riemann surface 
\refeq{lmir}. The boundary of the one-chain is specified by the 
fixed loci of the holomorphic involution $J$. In particular, 
if $J$ acts freely, the $\IR\IP^2$ free energy is 
zero. The relation between the superpotential and the chain 
period is straightforward for the resolved conifold geometry, 
but somewhat subtle in the present case. We will discuss 
this in detail below. 

Using the methods of \cite{AAHV}, we 
find that the  hypersurface \refeq{lmirt} 
admits a mirror involution acting as
\be\label{eq:mirinv}
(x,z,e^v,e^u)\rightarrow (-x,-ze^{-u},e^{-t_2-v},e^{-u}).
\ee
if the complex structure moduli satisfy $z_1=z_3$. 
This condition is related by mirror symmetry to the condition 
$t_1=t_3$ on the {\bf A}-model side, which follows from the action 
of $I$ on the K\"ahler class. 
The fixed point set has two components determined by  $v=t_2/2$ and 
respectively $v=t_2/2+i\pi$. 

In principle, the superpotential should be computed by integrating 
the holomorphic one-form $udv$ on a one-chain on the Riemann surface 
stretching between these two points. This is very similar to the 
computation of the superpotential for holomorphic branes \cite{GJS,LM,Mi}. 
Here we will mainly follow the approach of \cite{LM,Mi}. 
In that case one has an algebraic coordinate $z_0$ on the 
D-brane moduli space which can be in this case identified to $v$. 
The superpotential is a function $W(z_0,z_1,z_2,z_3)$ 
which has been shown in \cite{LM,Mi} to be a double logarithmic 
solution of a GKZ system. In order to obtain an expansion 
with correct integrality properties, we have to expand this function in
terms of flat coordinates $q_0,q_1,q_2,q_3$. $q_0$ is an open 
string flat coordinate associated to $z_0$, and $q_{1,2,3}$ are 
standard closed string flat coordinates. 

Our problem is mathematically very similar to the open string 
computation. At the classical level, we can formally 
think of the two orientifold 
planes as two holomorphic branes located at the points 
$z_0=\pm z_2^{1/2}$. We stress that this is just a formal analogy, 
motivated by the mathematical similarities between the two systems. 
We do not consider open string theories in this paper. 
However, this is the correct interpretation only at classical level. 
At quantum level, the correct coordinate on the D-brane moduli 
is the flat coordinate $q_0$. Therefore the positions of the 
two orientifold planes should be corrected to $q_0=\pm q_2^{1/2}$. 
Moreover, the condition $z_1=z_3$ translates 
into $q_1=q_3$. In conclusion, we claim that the exact $\IR\IP^2$ 
superpotential should be given by the following expression 
\be\label{eq:suppA} 
W_{\IR\IP^2}=W(q_2^{1/2}, q,q_2,q) - W(-q_2^{1/2}, q,q_2,q)
\ee
where $q=q_1=q_3$. 
Using the results of \cite{LM,Mi} we find 
\be\label{eq:suppB}
\begin{split}
W=\sum_{n_0,n_1,n_2,n_3}&\frac{z_0^{n_0}(-z_1)^{n_1}(-z_2)^{n_2}(-z_3)^{n_3}\Gamma(n_0)\Gamma(n_0-n_1+n_2-n_3)}
{\Gamma(1+n_0)\Gamma(1+n_1)\Gamma(1+n_3)\Gamma(1+n_0-3n_1+n_2)\Gamma(1+n_1-n_2+n_3)}\\
&\times\frac{1}{\Gamma(1+n_2-3n_3)}.
\end{split}
\ee
The open and closed flat coordinates are given by 
\be\label{opflat}
q_0=z_0e^{-f_1},~~
q_1=z_1e^{3f_1},~~q_2=z_2e^{-f_1-f_3},~~q_3=z_3e^{3f_3},
\ee
where
\be\label{flatfcns}
\begin{gathered}
f_1=\sum_{n_1,n_2,n_3}\frac{(-z_1)^{n_1}(-z_2)^{n_2}z_3^{n_3}\Gamma(3n_1-n_2)}{\Gamma(1+n_1)\Gamma(1+n_3)
\Gamma(1+n_1-n_2+n_3)^2\Gamma(1+n_2-3n_3)},\\
f_3=\sum_{n_1,n_2,n_3}\frac{z_1^{n_1}(-z_2)^{n_2}(-z_3)^{n_3}\Gamma(-n_2+3n_3)}{\Gamma(1+n_1)\Gamma(1+n_3)
\Gamma(1+n_1-n_2+n_3)^2\Gamma(1-3n_1+n_2)}.
\end{gathered}
\ee
In terms of flat coordinates, the superpotential has the following expansion 
\be\label{suppp}
\begin{split}
W=&q_0+\frac{1}{4}q_0^2-2q_0q_1+\frac{1}{9}q_0^3-q_0^2q_1+5q_0q_1^2+q_0q_1q_2+\frac{1}{16}q_0^4
-q_0^3q_1+\frac{7}{2}q_0^2q_1^2+q_0^2q_1q_2-32q_0q_1^3\\
&-4q_0q_1^2q_2-2q_0q_1q_2q_3+\frac{1}{25}q_0^5-q_0^4q_1+3q_0^3q_1^2+q_0^3q_1q_2-21q_0^2q_1^3
-3q_0^2q_1^2q_2-2q_0^2q_1q_2q_3\\
&-5489q_0q_1^4+35q_0q_1^3q_2+8q_0q_1^2q_2q_3+q_0q_1q_2^2q_3+5q_0q_1q_2q_3^2+\ldots 
-q_0^5q_1-\frac{164}{9}q_0^3q_1^3+\ldots~.
\end{split}
\ee
Therefore we obtain 
\be\label{unorsup}
W=2q_2^{\frac{1}{2}}-4q_1q_2^{\frac{1}{2}}+10q_1^2q_2^{\frac{1}{2}}-64q_1^3q_2^{\frac{1}{2}}+\ldots+\frac{2}{9}q_2^{\frac{3}{2}}
-6q_1^2q_2^{\frac{3}{2}}+\frac{536}{9}q_1^3q_2^{\frac{3}{2}}+\ldots+\frac{2}{25}q_2^{\frac{5}{2}}+\ldots~.
\ee
This expansion is in precise agreement with the $\IR\IP^2$ free energy 
computed in \refeq{freenC}. 

\section*{Acknowledgements}
We would like to thank Bobby Acharya and Harald Skarke for useful discussions and correspondence.

\appendix
\bigskip
\section{Edge Factors for Symmetric Maps} 

In this appendix we compute the edge factors for horizontal components 
$\Sigma_e$ which are preserved by the antiholomorphic involution
$\sigma$, that is $\tau_E(e)=e$. In this case $d_e$ must be odd, and 
$f_e:\Sigma_e\ra X$ is a $d_e:1$ cover of the $(-1,-1)$ curve
$C\subset X$. To simplify notation, throughout this section, 
we will drop the subscript $e$ from $\Sigma_e,f_e,d_e$. 
In terms of local coordinates, $f$ is given by 
$z=w^{d}$. Note that there is a induced torus action on the domain 
$\Sigma_e$ with weight $\lambda_w ={\lb_z\over d}$. The edge factor 
of this map is 
\be\label{eq:edgeA} 
C(e)={e_T\left(H^1(f^*(T_X))^{(\sigma,I),m}\right)
e_T\left(H^0(\Sigma,T_\Sigma)^{\sigma,m}\right)
\over e_T\left(H^0(f^*(T_X))^{(\sigma,I),m}\right)}.
\ee
The superscript $(\sigma, I)$ or $\sigma$ indicates that the we have 
to take the invariant part of the cohomology groups under the induced 
involution, as discussed in section three. The superscript $m$ denotes 
the moving part with respect to the torus action, as usual. 

A straightforward computation shows that the cohomology groups 
are given by 
\be\label{eq:edgeB} 
\begin{split}
&H^0(\Sigma, T_\Sigma)=\left\{(a_0+ a_1w +a_2w^2){\partial_w}\right\}\cr
& H^0(\Sigma, f^*T_X) =\left\{(b_0+b_1w+\ldots + b_{2d}w^{2d}){\partial_z} 
\right\}\cr
& H^1(\Sigma, f^*T_X)^{\vee} = \left\{(c_0+c_1w+ \ldots +c_{d-2})dw\otimes 
{\partial_u} +
(d_0+d_1w+\ldots+d_{d-2})dw\otimes {\partial_v}\right\}.\cr
\end{split}
\ee
In the last line we have used Kodaira-Serre duality, 
$H^1(\Sigma, f^*T_X)^{\vee}\simeq H^0(\Sigma, f^*(T_X^{\vee})\otimes 
\CO(K_\Sigma))$.  
The induced antiholomorphic involutions act as 
\be\label{eq:edgeC}
\begin{aligned}[b]
&\sigma:&  &a_n \ra (-1)^n{\overline a}_{2-n},& & n=0,1,2&\cr
&(\sigma,I):& &b_n\ra (-1)^n{\overline b}_{2d-n},& &n=0,\ldots,2d&\cr
&(\sigma,I):& &c_n\ra (-1)^{d-2-n}{\overline d}_{d-2-n},& &n=0,\ldots, d-2&\cr
&(\sigma,I):& &d_n\ra -(-1)^{d-2-n}{\overline c}_{d-2-n},&&n=0,\ldots, d-2.& 
\end{aligned}
\ee
Therefore the fixed subspaces are characterized by 
\be\label{eq:edgeD}
\begin{aligned}[b] 
& H^0(\Sigma, T_\Sigma)^{\sigma}:& &a_0={\overline a}_2,\ 
a_1=-{\overline a}_1& & &\cr
& H^0(\Sigma, f^*T_X)^{(\sigma,I)}:& &b_n= (-1)^n{\overline b}_{2d-n},& &n=0,\ldots,2d&\cr
& \left(H^1(\Sigma, f^*T_X)^{\vee}\right)^{(\sigma, I)}:& 
 &c_n= (-1)^{d-2-n}{\overline d}_{d-2-n},&  &n=0,\ldots, d-2.&\cr
\end{aligned}
\ee
Note that these real subspaces are not equipped with a canonical 
complex structure, but they are equipped with a canonical complex 
structure up to conjugation. Therefore the edge factor \refeq{edgeA}
has an ambiguity of order two which can be fixed by choosing a 
complex structure on the fixed subspaces. We do not know how these 
choices must be correlated from the first principles. 
Here we will make some arbitrary choices, and fix the signs 
during the computation as explained in section 3.1.
Then we obtain 
\be\label{eq:edgeE}
\begin{split}
&  H^0(\Sigma, T_\Sigma)^{\sigma,m}\simeq \left(-{\lambda_z\over d}\right)\cr
&  H^0(\Sigma, f^*T_X)^{(\sigma,I),m}\simeq 
(-\lb_z)\oplus \left(-{(d-1)\lb_z\over d}\right)\oplus\ldots 
\oplus\left(-{\lambda_z\over d}\right)\cr
&  H^1(\Sigma, f^*T_X)^{(\sigma,I),m}\simeq
\left(-\lb_v-{\lb_z\over d}\right)\oplus 
\left(-\lb_v-{2\lb_z\over d}\right)\oplus \ldots 
\left(-\lb_v-{(d-1)\lb_z\over d}\right).\cr
\end{split}
\ee
By substituting this back in equation \refeq{edgeA}, we obtain 
formula \refeq{locG} in the main text.

\renewcommand{\baselinestretch}{1.0} \normalsize


\bibliography{strings,m-theory,susy,largeN}

\providecommand{\href}[2]{#2}\begingroup\raggedright\begin{thebibliography}{10}

\bibitem{AAHV}
B.~Acharya, M.~Aganagic, K.~Hori and C.~Vafa,
``Orientifolds, Mirror Symmetry and Superpotentials'',  
\href{http://xxx.lanl.gov/abs/hep-th/0202208}
{{\tt hep-th/0202208}}.

\bibitem{Alling} 
N.L.~Alling, {\it Foundations of The Theory of Klein Surfaces}, 
Lecture Notes in Mathematics {\bf 219}, Spinger-Verlag, 1971. 

\bibitem{AV:mirror}
M.~Aganagic and C.~Vafa, ``Mirror
Symmetry, D-Branes and Counting Holomorphic Discs'', 
\href{http://xxx.lanl.gov/abs/hep-th/0202208}
{{\tt hep-th/0012041}}.

\bibitem{AKV:disk}
M.~Aganagic, A.~Klemm and C.~Vafa, ``Disk Instantons, 
Mirror Symmetry and the Duality Web'',
Z. Naturforsch. {\bf A 57} (2002) 1, 
\href{http://xxx.lanl.gov/abs/hep-th/0105045}
{{\tt hep-th/0105045}}. 

\bibitem{B}
K. Behrend, ``Gromov-Witten Invariants in Algebraic Geometry'', 
Invent. Math. {\bf 127} (1997) 601.

\bibitem{BF}
K. Behrend and B. Fantechi, ``The
Intrinsic Normal Cone'', Invent. Math. {\bf 128} (1997) 45.

\bibitem{BH} 
I. Brunner and K. Hori, ``Orientifolds and Mirror Symmetry'', 
\href{http://xxx.lanl.gov/abs/hep-th/0303135}
{{\tt hep-th/0303135}}.

\bibitem{CKYZ}
T.-M. Chiang, A. Klemm, S.-T. Yau and E. Zaslow, ``Local Mirror
Symmetry: Calculations and Interpretations'', ATMP {\bf 3} (1999)
495, \href{http://xxx.lanl.gov/abs/hep-th/9903053}{{\tt hep-th/9903053}}.

\bibitem{DM}
P. Deligne and D. Mumford, ``The Irreducibility of The Moduli 
Space of Curves of Given Genus'', Inst. Hautes {\' E}tudes Sci. Publ. 
Math. {\bf 36} (1069) 75. 


\bibitem{DVi:matrix} 
R.~Dijkgraaf and C.~Vafa, ``Matrix Models, Topological Strings, and 
Supersymmetric Gauge Theories'', 
\href{http://xxx.lanl.gov/abs/hep-th/0206255}{{\tt hep-th/0206255}}.

\bibitem{DVii:geom}
R.~Dijkgraaf and C.~Vafa, ``On Geometry and Matrix Models'', 
\href{http://xxx.lanl.gov/abs/hep-th/0207106}{{\tt hep-th/0207106}}.

\bibitem{FP}
C. Faber and R. Pandharipande, ``Hodge Integrals and
Gromov--Witten Theory'', Invent. Math. {\bf 139} (2000) 139, \href{http://xxx.lanl.gov/abs/math.AG/9810173}
{{\tt math.AG/9810173}}.

\bibitem{GJS}
S. Govindarajan, T. Jayaraman and T. Sarkar, ``Disc
Instantons in Linear Sigma Models'', \href{http://xxx.lanl.gov/abs/hep-th/0108234}
{{\tt hep-th/0108234}}.

\bibitem{GP}
T. Graber and R. Pandharipande,
``Localization of Virtual Classes'', Invent. Math. {\bf 135}
(1999) 487, \href{http://xxx.lanl.gov/abs/math.AG/9708001}
{{\tt math.AG/9708001}}.

\bibitem{GZ}
T. Graber and E. Zaslow, ``Open String Gromov-Witten
Invariants: Calculations and a Mirror 'Theorem' '', 
\href{http://xxx.lanl.gov/abs/hep-th/0109075}
{{\tt hep-th/0109075}}.

\bibitem{Hi}
P. Ho${\check{\rm{r}}}$ava, ``Strings on World-Sheet Orbifolds'', 
Nucl. Phys. {\bf B327} (1989) 461.

\bibitem{Hii} 
P. Ho${\check{\rm{r}}}$ava, ``Equivariant Topological Sigma Models'', 
Nucl. Phys. {\bf B418} (1994) 571, 
\href{http://xxx.lanl.gov/abs/hep-th/9309124}
{{\tt hep-th/9309124}}.

\bibitem{Hiii} 
P. Ho${\check{\rm{r}}}$ava, 
``Chern-Simons Gauge Theory on Orbifolds: Open Strings from Three Dimensions'',
J. Geom. Phys. {\bf 21} (1996) 1, 
\href{http://xxx.lanl.gov/abs/hep-th/9404101}
{{\tt hep-th/9404101}}.

\bibitem{HV}
K. Hori and C. Vafa, ``Mirror Symmetry'', 
\href{http://xxx.lanl.gov/abs/hep-th/0002222}
{{\tt hep-th/0002222}}.

\bibitem{KL}
S. Katz and C.-C. M. Liu, ``Enumerative Geometry of Stable
Maps with Lagrangian Boundary Conditions and Multiple Covers of
the Disc'', ATMP {\bf 5} (2001) 1, \href{http://xxx.lanl.gov/abs/math.AG/0103074}
{{\tt math.AG/0103074}}.

\bibitem{FKi} 
F. Klein, ``\"Uber eine neue Art von Riemannschen Fl\"achen'', 
Math. Ann. {\bf 10} (1876).

\bibitem{FKii} 
F. Klein, ``\"Uber Realit\"atsverh\"altnisse bei der einem beliebigen 
Geschlechte zugeh\"origen Normalkurve
der $\phi$'', Math. Ann. {\bf 42} (1892).

\bibitem{K}
M. Kontsevich, ``Enumeration of Rational Curves via Torus Actions'',
{\it The Moduli Space of Curves}, 335-368, Progr. Math. {\bf 129},
Birkh\"auser Boston, MA, 1995.

\bibitem{KM}
M. Kontsevich and Yu.I. Manin, ``Gromov-Witten Classes, Quantum Cohomology, and Enumerative Geometry'', 
Commun. Math. Phys. {\bf 164} (1994) 525, 
\href{http://xxx.lanl.gov/abs/hep-th/9402147}
{{\tt hep-th/9402147}}.

\bibitem{LM}
W. Lerche and P. Mayr, ``On ${\cal N}=1$ Mirror Symmetry
for Open Type II Strings'', \href{http://xxx.lanl.gov/abs/hep-th/0111113}
{{\tt hep-th/0111113}}.

\bibitem{LMWi}
W. Lerche, P. Mayr and N. Warner, 
`` Holomorphic ${\cal N}=1$ Special Geometry of Open--Closed Type II Strings'', 
\href{http://xxx.lanl.gov/abs/hep-th/0207259}{{\tt hep-th/0207259}}.

\bibitem{LMWii}
W. Lerche, P. Mayr and N. Warner,
``${\cal N}=1$ Special Geometry, Mixed Hodge Variations and Toric Geometry'', 
\href{http://xxx.lanl.gov/abs/hep-th/0208039}{{\tt hep-th/0208039}}.

\bibitem{LS}
J. Li and Y.S. Song, ``Open String Instantons and Relative Stable
Morphisms'', Adv. Theor. Math. Phys. {\bf 5} (2002) 67, \href{http://xxx.lanl.gov/abs/hep-th/0103100}
{{\tt hep-th/0103100}}.

\bibitem{LT}
J. Li~and G. Tian,
``Virtual Moduli Cycles and Gromov-Witten Invariants of Algebraic Varieties'',
J. Amer. Math. Soc. {\bf 11} (1998) 119.

\bibitem{YM}
Yu.I. Manin, ``Generating Functionals in Algebraic Geometry and Sums 
over Trees'', in {\it The Moduli Space of Curves}, Progr. Math. {\bf 129},
Birkh\"auser, Boston-Basel-Berlin, 1995.

\bibitem{Mi}
P. Mayr, ``${\cal N}=1$ Mirror Symmetry and Open/Closed String
Duality'', \href{http://xxx.lanl.gov/abs/hep-th/0108229}
{{\tt hep-th/0108229}}.             

\bibitem{Mii}
P. Mayr, 
``Summing up Open String Instantons and ${\cal N}=1$ String Amplitudes'', 
\href{http://xxx.lanl.gov/abs/hep-th/0203237}
{{\tt hep-th/0203237}}. 

\bibitem{AM1} A. Misra, ``An ${\cal N}=1$ Triality by Spectrum Matching", {{\tt hep-th/0212054}}.

\bibitem{AM2} A. Misra, ``On (Orientifold of) Type IIA on a Compact Calabi-Yau", 
{{\tt hep-th/0304209}}.

\bibitem{OV:knot}
H.~Ooguri and C. Vafa,
``Knot Invariants and Topological Strings'', 
Nucl. Phys. {\bf B 577} (2000) 419-438,  
\href{http://xxx.lanl.gov/abs/hep-th/9912123}
{{\tt hep-th/9912123}}.

\bibitem{SV}
S. Sinha and C. Vafa, ``$SO$ and $Sp$ at Large $N$, \href{http://xxx.lanl.gov/abs/hep-th/0012136}
{{\tt hep-th/0012136}}.

\bibitem{VW}
C. Vafa and E. Witten, ``Dual String Pairs with ${\cal N}=1$ and ${\cal N}=2$ Supersymmetry in Four Dimensions'', 
Nucl. Phys. Proc. Suppl. {\bf 46} (1996) 225, \href{http://xxx.lanl.gov/abs/hep-th/9507050}
{{\tt hep-th/9507050}}.

\bibitem{EW}
E. Witten, ``Two-Dimensional Gravity and Intersection Theory on 
The Moduli Space'', {\it Surveys in Differential Geometry}, {\bf 1} 
(1991) 243. 


\end{thebibliography}\endgroup
\bibliographystyle{utphys}

\end{document}